# Measurement and Modeling of Polarized Atmosphere at the South Pole with SPT-3G

A. Coerver[1], J. A. Zebrowski[2,3,4,40], S. Takakura[5,6,7], W. L. Holzapfel[1], P. A. R. Ade[8], A. J. Anderson[2,3,4], Z. Ahmed[9,10,11], B. Ansarinejad[12], M. Archipley[2,3], L. Balkenhol[13], D. Barron[14], K. Benabed[13], A. N. Bender[2,3,15], B. A. Benson[2,3,4], F. Bianchini[9,10,11], L. E. Bleem[2,15], F. R. Bouchet[13], L. Bryant[16], E. Camphuis[13], J. E. Carlstrom[2,3,15,16,17], T. W. Cecil[15], C. L. Chang[2,3,15], P. Chaubal[12], P. M. Chichura[2,17], A. Chokshi[18], T.-L. Chou[2,17], T. M. Crawford[2,3], A. Cukierman[9,10,11], C. Daley[19], T. de Haan[20], K. R. Dibert[2,3], M. A. Dobbs[21,22], A. Doussot[13], D. Dutcher[23], W. Everett[24], C. Feng[25], K. R. Ferguson[26,27], K. Fichman[2,17], A. Foster[23], S. Galli[13], A. E. Gambrel[2], R. W. Gardner[16], F. Ge[28], N. Goeckner-Wald[9,11], R. Gualtieri[29], F. Guidi[13], S. Guns[1], N. W. Halverson[7,30], E. Hivon[13], G. P. Holder[25], J. C. Hood[2], A. Hryciuk[2,17], N. Huang[1], F. Kéruzoré[15], A. R. Khalife[13], L. Knox[28], M. Korman[31], K. Kornoelje[2,3], C.-L. Kuo[9,10,11], A. T. Lee[1,32], K. Levy[12], A. E. Lowitz[2], C. Lu[25], A. Maniyar[9,10,11], E. S. Martsen[2,3], F. Menanteau[19,33], M. Millea[1], J. Montgomery[21], Y. Nakato[11], T. Natoli[2], G. I. Noble[34,35], V. Novosad[36], Y. Omori[3,2], S. Padin[2,37], Z. Pan[2,15,17], P. Paschos[16], K. A. Phadke[19,33], A. W. Pollak[18], K. Prabhu[28], W. Quan[2,17], M. Rahimi[12], A. Rahlin[2,4], C. L. Reichardt[12], M. Rouble[21], J. E. Ruhl[31], E. Schiappucci[12], G. Smecher[38], J. A. Sobrin[2,4], A. A. Stark[39], J. Stephen[16], A. Suzuki[32], C. Tandoi[19], K. L. Thompson[9,10,11], B. Thorne[28], C. Trendafilova[33], C. Tucker[8], C. Umilta[25], J. D. Vieira[19,25,33], A. Vitrier[13], Y. Wan[19,33], G. Wang[15], N. Whitehorn[27], W. L. K. Wu[9,10], V. Yefremenko[15], and M. R. Young[2,4]

[1] Department of Physics, University of California, Berkeley, CA 94720, USA; acoerver@berkeley.edu
[2] Kavli Institute for Cosmological Physics, University of Chicago, 5640 South Ellis Avenue, Chicago, IL 60637, USA
[3] Department of Astronomy and Astrophysics, University of Chicago, 5640 South Ellis Avenue, Chicago, IL 60637, USA
[4] Fermi National Accelerator Laboratory, MS209, P.O. Box 500, Batavia, IL 60510, USA
[5] International Center for Quantum-field Measurement Systems for Studies of the Universe and Particles (QUP), High Energy Accelerator Research Organization (KEK), Tsukuba, Ibaraki 305-0801, Japan
[6] Department of Physics, The University of Tokyo, Tokyo 113-0033, Japan
[7] CASA, Department of Astrophysical and Planetary Sciences, University of Colorado, Boulder, CO 80309, USA
[8] School of Physics and Astronomy, Cardiff University, Cardiff CF24 3YB, UK
[9] Kavli Institute for Particle Astrophysics and Cosmology, Stanford University, 452 Lomita Mall, Stanford, CA 94305, USA
[10] SLAC National Accelerator Laboratory, 2575 Sand Hill Road, Menlo Park, CA 94025, USA
[11] Department of Physics, Stanford University, 382 Via Pueblo Mall, Stanford, CA 94305, USA
[12] School of Physics, University of Melbourne, Parkville, VIC 3010, Australia
[13] Sorbonne Universit'e, CNRS, UMR 7095, Institut d'Astrophysique de Paris, 98 bis bd Arago, 75014 Paris, France
[14] Department of Physics and Astronomy, The University of New Mexico, Albuquerque, NM 87131, USA
[15] High-Energy Physics Division, Argonne National Laboratory, 9700 South Cass Avenue., Lemont, IL 60439, USA
[16] Enrico Fermi Institute, University of Chicago, 5640 South Ellis Avenue, Chicago, IL 60637, USA
[17] Department of Physics, University of Chicago, 5640 South Ellis Avenue, Chicago, IL 60637, USA
[18] University of Chicago, 5640 South Ellis Avenue, Chicago, IL 60637, USA
[19] Department of Astronomy, University of Illinois Urbana-Champaign, 1002 West Green Street, Urbana, IL 61801, USA
[20] High Energy Accelerator Research Organization (KEK), Tsukuba, Ibaraki 305-0801, Japan
[21] Department of Physics and McGill Space Institute, McGill University, 3600 Rue University, Montreal, Quebec H3A 2T8, Canada
[22] Canadian Institute for Advanced Research, CIFAR Program in Gravity and the Extreme Universe, Toronto, ON, M5G 1Z8, Canada
[23] Joseph Henry Laboratories of Physics, Jadwin Hall, Princeton University, Princeton, NJ 08544, USA
[24] Department of Astrophysical and Planetary Sciences, University of Colorado, Boulder, CO 80309, USA
[25] Department of Physics, University of Illinois Urbana-Champaign, 1110 West Green Street, Urbana, IL 61801, USA
[26] Department of Physics and Astronomy, University of California, Los Angeles, CA 90095, USA
[27] Department of Physics and Astronomy, Michigan State University, East Lansing, MI 48824, USA
[28] Department of Physics & Astronomy, University of California, One Shields Avenue, Davis, CA 95616, USA
[29] Department of Physics and Astronomy, Northwestern University, 633 Clark Street, Evanston, IL 60208, USA
[30] Department of Physics, University of Colorado, Boulder, CO 80309, USA
[31] Department of Physics, Case Western Reserve University, Cleveland, OH 44106, USA
[32] Physics Division, Lawrence Berkeley National Laboratory, Berkeley, CA 94720, USA
[33] Center for AstroPhysical Surveys, National Center for Supercomputing Applications, Urbana, IL 61801, USA
[34] Dunlap Institute for Astronomy & Astrophysics, University of Toronto, 50 St. George Street, Toronto, ON, M5S 3H4, Canada
[35] David A. Dunlap Department of Astronomy & Astrophysics, University of Toronto, 50 St. George Street, Toronto, ON, M5S 3H4, Canada
[36] Materials Sciences Division, Argonne National Laboratory, 9700 South Cass Avenue, Lemont, IL 60439, USA
[37] California Institute of Technology, 1200 East California Boulevard., Pasadena, CA 91125, USA
[38] Three-Speed Logic, Inc., Victoria, B.C., V8S 3Z5, Canada
[39] Harvard-Smithsonian Center for Astrophysics, 60 Garden Street, Cambridge, MA 02138, USA

---

[40] NASA Einstein Fellow.

## Abstract

We present the detection and characterization of fluctuations in linearly polarized emission from the atmosphere above the South Pole. These measurements make use of data from the SPT-3G receiver on the South Pole Telescope in three frequency bands centered at 95, 150, and 220 GHz. We use the cross-correlation between detectors to produce an unbiased estimate of the power in Stokes $I$, $Q$, and $U$ parameters on large angular scales. Our results are consistent with the polarized signal being produced by the combination of Rayleigh scattering of thermal radiation from the ground and thermal emission from a population of horizontally aligned ice crystals with an anisotropic distribution described by Kolmogorov turbulence. The measured spatial scaling, frequency scaling, and elevation dependence of the polarized emission are explained by this model. Polarized atmospheric emission has the potential to significantly impact observations on the large angular scales being targeted by searches for inflationary B-mode CMB polarization. We present the distribution of measured angular power spectrum amplitudes in Stokes $Q$ and $I$ for 4 yr of Austral winter observations, which can be used to simulate the impact of atmospheric polarization and intensity fluctuations at the South Pole on a specified experiment and observation strategy. We present a mitigation strategy that involves both downweighting significantly contaminated observations and subtracting a polarized atmospheric signal from the 150 GHz band maps. In observations with the SPT-3G instrument, the polarized atmospheric signal is a well-understood and subdominant contribution to the measured noise after implementing the mitigation strategies described here.

*Unified Astronomy Thesaurus concepts:* Cosmology (343); Atmospheric effects (113); Cosmic microwave background radiation (322); Millimeter astronomy (1061)

## 1. Introduction

Precision measurements of the cosmic microwave background (CMB) temperature and polarization anisotropy serve as a cornerstone of modern cosmology. The search for degree-scale odd-parity (B-mode) polarization arising from gravitational waves produced in the inflationary epoch (U. Seljak & M. Zaldarriaga 1997) is the primary scientific focus of many current and planned CMB experiments. Predictions for the amplitude of this signal are uncertain; however, current experiments limit it to be <10 nK on the degree angular scales where it is predicted to peak (P. A. R. Ade et al. 2021). The detection of this signal is challenging due to the extreme instrument sensitivity and control of systematic errors required. In particular, a robust detection of the inflationary B-mode signal will require careful control of astrophysical foregrounds and sensitive measurements over a broad range of angular scales and frequencies. Ground-based observations of the CMB face the additional challenge of emission from the atmosphere. In the millimeter-wavelength bands typically used for ground-based CMB observations, atmospheric emission is dominated by the wings of oxygen and water lines and can be considered optically thin with an opacity of a few percent (J. Pardo et al. 2001). This contributes a constant power loading on the detectors that increases their fundamental noise set by the statistical arrival of photons. The distribution of water vapor is anisotropic, and fluctuations in the intensity of the emitted radiation result in an additional source of noise (O. P. Lay & N. W. Halverson 2000). For these reasons, ground-based CMB experiments are typically placed at high-altitude sites where the atmosphere is particularly thin and dry, such as the South Pole and the Atacama Desert in the Chilean Andes.

Ground-based CMB experiments optimized for polarization measurements are designed to reject atmospheric temperature fluctuations and recover their full sensitivity to CMB polarization anisotropy. Forecasts for the performance of future CMB experiments typically assume that atmospheric emission is entirely unpolarized. However, some simulations have considered the impact of polarized atmosphere (L. Pietranera et al. 2007). Recently, two CMB experiments published detections of linearly polarized atmospheric emission originating from horizontally aligned ice crystals in the atmosphere (S. Takakura et al. 2019; Y. Li et al. 2023). Another potential source of polarized emission is circular polarization from oxygen molecules due to Zeeman splitting by the Earth's magnetic field, which is discussed in B. Keating et al. (1998), S. Hanany & P. Rosenkranz (2003), S. Spinelli et al. (2011), and measured by the CLASS experiment (M. A. Petroff et al. 2020). SPT-3G is, by design, not sensitive to circular polarization. That, in combination with the uniform distribution of oxygen in the atmosphere, means that circularly polarized atmospheric emission is not a source of noise for measurements with SPT-3G.

In this work, we present a significant detection of linearly polarized atmospheric emission at the South Pole with SPT-3G. We show that this polarized atmospheric emission is caused by thermal emission and scattering of radiation from the ground by horizontally aligned ice crystals as they fall under the influence of gravity. Due to the horizontal alignment of the ice crystals, the atmospheric signal, shown schematically in Figure 1, is horizontally polarized. Polarization of the atmospheric signal has the potential to significantly impact the ability of ground-based telescopes to measure large-scale CMB polarization. To address this, it is essential to characterize the atmospheric polarization fluctuations at the sites of current and proposed CMB observatories.

This paper provides the first detection and detailed characterization of polarized emission from the atmosphere above the South Pole, as well as the full distribution of polarized emission fluctuation amplitudes during the Austral winter. We use data from the SPT-3G receiver on the South Pole Telescope (SPT) in three bands centered at 95, 150, and 220 GHz during the 2019–2022 Austral winter observing seasons. The measurements are based on cross correlations between independent detectors in order to remove noise bias and maximize the sensitivity to polarized emission. We characterize the amplitude, polarization angle, and spectral and spatial scaling of the polarized signal and show that the results are consistent with the expectations of scattering and emission from an anisotropic distribution of ice crystals in the atmosphere.

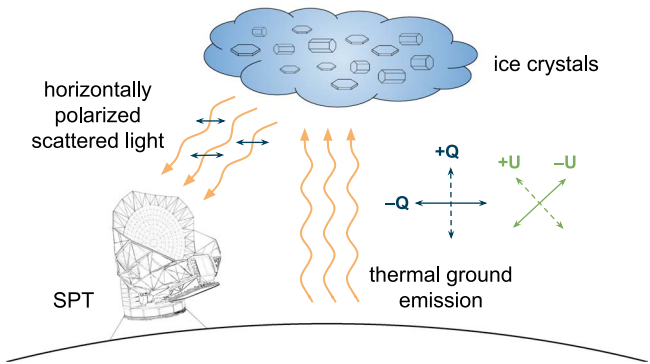

**Figure 1.** This Figure illustrates radiation from the ground being scattered by horizontally aligned ice crystals. This results in horizontally polarized (negative $Q$, zero $U$ Stokes parameters) radiation detected at the telescope.

The results of this paper can be used to simulate the impact of polarized atmospheric emission for observations from the South Pole given the details of the experiment, observation strategy, and analysis pipeline. This is particularly important for producing accurate forecasts, optimizing instrument configurations, and planning observation strategies for future experiments probing large-scale polarization, such as CMB-S4 (K. Abazajian et al. 2022).

This paper is organized as follows: In Section 2, we provide background and describe progress in the measurement of atmospheric temperature and polarization anisotropy at millimeter wavelengths. In Section 3, we present a comprehensive theoretical model for the polarized signal arising from ice crystals in the atmosphere. Section 4 describes the spatial modeling of the signal and shows how 1D telescope scans are used to measure the amplitude and spatial scaling of the polarized signal. In Section 5, we discuss the SPT-3G data set and the processing of the data. Section 6 discusses the measurement of the wind speed and the applicability of the frozen sky approximation. In Section 7, we present measurements of the spatial and spectral scaling and amplitude distribution of the temperature anisotropy power. In Section 8, we present the measurements of the amplitude, polarization angle, and spectral and spatial scaling of the polarization anisotropy power. We show that both temperature and polarization anisotropy power are consistent with theoretical models. Section 9 presents a prescription for using the results of this work to simulate realizations of atmospheric temperature and polarization fluctuations. In Section 10, we describe a set of methods that use the temporal variability and frequency scaling of the polarized emission to mitigate its impact on the SPT-3G large-scale polarization maps. We summarize and present conclusions in Section 11.

## 2. Anisotropic Atmospheric Emission

### 2.1. Unpolarized Atmospheric Emission

The main sources of emission from the atmosphere at millimeter wavelengths are the vibrational/rotational transitions of atmospheric water vapor and oxygen. The $O_2$ molecule has strong absorption/emission lines near 60 and at 119 GHz, while $H_2O$ vapor has absorption/emission lines at 22 and 180 GHz. The frequency bands of ground-based CMB experiments are chosen to avoid the centers of these lines. However, emission from the broad wings of oxygen and water vapor lines still produce the majority of atmospheric emission in CMB frequency bands. Water vapor dominates the atmospheric absorption and emission at frequencies above 120 GHz, where the brightness of the CMB peaks and most CMB experiments have observing bands.

The fundamental sensitivity reached by the detectors in a CMB experiment is limited by the statistical arrival of photons (J. Zmuidzinas 2003; C. A. Hill & A. Kusaka 2024). A higher background power increases this photon noise and reduces the experiment sensitivity. In a well-designed experiment, radiation from the atmosphere can dominate the total power reaching the detectors. The emission from water vapor is highly variable in time and scales with the precipitable water vapor (PWV), which is the equivalent thickness of liquid water in a vertical column above the observation site. This emission is minimized at high-altitude sites where the atmosphere is thinner and drier. For this reason, many CMB experiments deploy telescopes at the driest sites in the world, such as the South Pole and the Atacama Desert.

Unlike oxygen, which is uniformly distributed in the atmosphere, water vapor is anisotropically distributed and, in addition to increasing photon noise, produces fluctuations in emission that add noise to measurements of CMB temperature anisotropy. The spatial distribution of the water vapor is described by the Kolmogorov theory of turbulence, where energy is input by shear on larger scales and then cascades to smaller scales where it eventually dissipates (A. Kolmogorov 1941). In this theory, the fluctuation power in a large 3D volume will scale as a function of spatial wavenumber, $k$, as $P(k) \propto k^{-11/3}$. V. I. Tatarskii (1961) showed that this same power-law scaling applies to the distribution of constituents that are passively entrained in the turbulence, such as water vapor or ice crystals. When observed from the ground, the atmosphere can, to a good approximation, be modeled as a 2D screen with power that decreases rapidly with increasing angular wavenumber, $\alpha$, as $P(\alpha) \propto \alpha^{-8/3}$ (R. S. Bussmann et al. 2005). In some previous work, the structure and time evolution of the full 3D power distribution has been modeled (S. E. Church 1995; J. Errard et al. 2015; T. W. Morris et al. 2022). However, the impact of that more complex modeling on the derived parameters of the atmosphere is negligible compared to other uncertainties and approximations (J. Errard et al. 2015).

The signal due to atmospheric intensity fluctuations has been evaluated in depth by O. P. Lay & N. W. Halverson (2000) for a wide variety of experimental configurations and observing strategies. On large angular scales, these atmospheric fluctuations are much brighter than the CMB anisotropy that we seek to measure. Fortunately, the atmosphere changes in time, and this signal becomes a source of noise that can be averaged down over many observations. However, atmospheric fluctuation power remains the dominant source of low-frequency noise for ground-based observations of CMB temperature anisotropies. Emission from atmospheric water vapor is not expected to be polarized; however, leakage of temperature to polarization signals caused by instrumental effects such as polarization by reflections, detector pair gain mismatch, or systematic errors introduced by polarization modulators can lead to additional noise in polarization.

The geographic South Pole is the premier site for millimeter-wave measurements of the CMB on large angular scales due to its high altitude, low atmospheric water vapor, stable thermal environment during the Austral winter, and established infrastructure to support Austral winter observing. The median PWV at the South Pole is 0.32 mm (C.-L. Kuo 2017; H. Yang et al. 2010), a factor of 3 lower than at the Chajnantor Science Reserve in the Atacama Desert where the CLASS (T. Essinger-Hileman et al. 2014), ACT (D. S. Swetz et al. 2011), POLARBEAR (Z. D. Kermish et al. 2012), and Simons Observatory (N. Galitzki et al. 2018) experiments are located. Perhaps more importantly, the median variability in the PWV is more than a factor of 10 lower at the South Pole than at the Chajnantor site (C.-L. Kuo 2017). R. S. Bussmann et al. (2005) showed that the measured atmospheric intensity fluctuations at the South Pole were consistent with an anisotropic distribution of water vapor described by Kolmogorov turbulence. They solved for the amplitude of the atmospheric power fluctuations in observation bands centered at 150, 220, and 274 GHz for each 2 hr observation period over the course of an entire 6 month Austral winter observing season. The spatial scaling of the atmospheric fluctuation power above the Chajnantor site was measured by the ACT experiment (T. W. Morris et al. 2022) and was also shown to be consistent with the predictions of Kolmogorov turbulence. The temperature anisotropy above the Chajnantor site was characterized by J. Errard et al. (2015) with the POLARBEAR experiment, who found the median fluctuation power at 150 GHz to be ∼100 times that measured by R. S. Bussmann et al. (2005) above the South Pole.

## 2.2. Polarized Atmospheric Emission

Less is known about fluctuations in polarized atmospheric emission due to the much smaller amplitude. In this work, we find that radiation scattered and emitted by ice crystals is the dominant source of millimeter-wavelength atmospheric linear polarization above the South Pole.

The potential impact of atmospheric ice crystals on CMB polarization measurements was pointed out in L. Pietranera et al. (2007). The POLARBEAR experiment detected bursts of horizontally polarized signal, which coincided with the appearance of clouds in an optical camera (S. Takakura et al. 2019). At 150 GHz, some of these bursts had amplitude $\Delta|Q| > 0.3$ $K_{RJ}$, where $K_{RJ}$ denotes Rayleigh–Jeans temperature units in kelvin. Recently, the CLASS experiment was used to measure atmospheric polarization in bands centered at 40, 90, 150, and 220 GHz (Y. Li et al. 2023). High signal-to-noise detections of polarized emission were found to coincide with the appearance of clouds above the observing site. The polarized signal was detected in all four frequency bands and was horizontally polarized, as is expected from gravitationally aligned ice crystals. At 220 GHz, the largest observed fluctuations had amplitudes $\Delta|Q| > 1.0$ $K_{RJ}$. From the relative power in the 90 and 150 GHz bands, they found the spectral index of the polarized emission to be $\alpha = 3.90 \pm 0.06$, consistent with the Rayleigh scattering of thermal radiation from the ground by horizontally aligned ice crystals. However, the spectral index from all four bands was found to be $\alpha = 3.17 \pm 0.05$ and deviated from a single power-law scaling at 220 GHz. They interpret this as being potentially due to ice crystals sufficiently large that Mie scattering rather than Rayleigh scattering is appropriate. This explanation is consistent with the observed extremely large polarized signal.

The BICEP/Keck project has also reported evidence of excess correlated polarized noise, which could be interpreted as polarized atmospheric fluctuations above the South Pole (B. Singari & BICEP/Keck Team 2023). This manifests as excess large angular scale noise in a fraction of their observations. In this work, we characterize not just episodes of intense polarized emission, but the complete distribution of polarized atmospheric fluctuation power seen above the South Pole over 4 yr of observation with SPT-3G.

### 2.2.1. Morphology of Ice Crystals at the South Pole

Ice crystals widely exist in the atmosphere in the form of cirrus clouds and precipitation. At the South Pole, they are commonly found in the first 3 km above the ground (R. P. Lawson et al. 2006). Ice crystals exhibit large variations in size and shape depending on the location and atmospheric conditions. There have been several measurements of the properties of ice crystals in the atmosphere above the South Pole. R. P. Lawson et al. (2006) measured falling ice crystals in the Austral summer for 9 days and characterized their sizes and shapes. In a later study, also during the Austral summer, R. P. Lawson et al. (2011) used in situ measurements from a tethered balloon to measure ice crystal properties at the South Pole. In the most relevant study for this work, V. P. Walden et al. (2003) collected and measured falling ice crystals at the South Pole during the Austral winter. They divided the observed ice crystals into three main morphological groups: "diamond dust," "blowing snow," and "snow grains." Diamond dust consists primarily of relatively small hexagonal columns and plates. The crystal length ($c$) ranges from 3–1000 $\mu$m, and the width ($2a$) ranges from 2–158 $\mu$m. The mode of the aspect ratio ($c/2a$) is 4 (0.5) for columns (plates), and equidimensional crystals ($c \approx 2a$) are rare. The median equivalent radius of these particles is $r_e \sim 12$ $\mu$m with an upper limit of $r_e < 30$ $\mu$m. Ice crystal size is correlated with the temperature of the regions where they are formed (R. T. Austin et al. 2009). The typical low temperature and humidity of the atmosphere above the South Pole are responsible for the slow growth and small size of these crystals.

Blowing snow ice crystals are typically small, round, and located close to the ground. They have a median equivalent radius of $r_e \sim 11$ $\mu$m with an upper limit of $r_e < 30$ $\mu$m. Blowing snow is driven by wind and is often present at the snow surface, but is relegated to a layer of at most tens of meters above the surface.

About 7% of samples collected by V. P. Walden et al. (2003) included crystals characterized as snow grains, which had precipitated from clouds. These large ice crystals are rare compared to the smaller "diamond dust" crystals, but contain approximately half of the total collected ice volume. The largest of these are described as "bullet clusters" (or "rosettes" in the terminology of R. P. Lawson et al. 2006), which are aggregates of hollow hexagonal crystals and large solid hexagonal columns. Both of these crystal types have a broad distribution of sizes with an upper limit near an equivalent radius of 100 $\mu$m. These crystals are believed to form when a supercooled water drop freezes rapidly (R. P. Lawson et al. 2006). We will argue in Section 3 that these rare and large ice crystals are likely responsible for the majority of the observed polarized signal.

Ice crystal alignment is dependent on particle size and strength of atmospheric turbulence. In a calm atmosphere, horizontal alignment is expected for both plates and columns. In a moderate to strongly turbulent atmosphere, columns with a radius of $\lesssim 5$ $\mu$m and plates with a radius of $\lesssim 10$ $\mu$m can become misaligned (J. D. Klett 1995). Average tilt angle decreases as crystal size increases, resulting in an average tilt angle of $<3°$ above a radius of $50\,\mu m$ for both plates and columns. Exact horizontal alignment of individual ice crystals is not required to produce a horizontal polarization signal. All that is required is that the population of ice crystals is preferentially horizontally aligned.

## 3. Signal from Ice Crystals at Millimeter Wavelengths

In this Section, we describe the unpolarized and polarized signal at millimeter wavelengths expected from the combination of scattering and emission by ice crystals in the atmosphere. Most significantly, we derive expressions for the frequency scaling of the power and its dependence on the elevation angle of the telescope.

### 3.1. Unpolarized Signals from Ice Crystals

Details of the scattering theory of millimeter waves by small ice crystals can be found in textbooks (e.g., L. D. Landau & E. M. Lifshitz 1960; C. F. Bohren & D. R. Huffman 1998; M. I. Mishchenko et al. 2002; C. Mätzler 2006). Following the treatment of S. Takakura et al. (2019), we assume the small particle limit. First, we consider spherical particles to estimate the contributions from scattering and emission. Then, we consider the nonspherical shape of the ice crystals, which is responsible for the majority of the polarized signal.

The cross sections for scattering and absorption can be written as

$$\sigma_{sca} = \frac{1}{6\pi}\frac{\omega^4}{c^4}V^2|A|^2, \quad (1)$$

$$\sigma_{abs} = \frac{\omega}{c}V \;\; \text{Im}[A], \quad (2)$$

where $\omega = 2\pi\nu$ is the angular frequency of the radiation, $V$ is the volume of the particle, and $c$ is the speed of light. The effects of the dielectric properties of the ice and particle shape are included in the polarizability $A$. In the case of spherical particles, this becomes $A = 3(\varepsilon - 1)/(\varepsilon + 2)$, where $\varepsilon = \varepsilon' + i\varepsilon''$ is the complex relative permittivity of ice. Following the model of S. G. Warren & R. E. Brandt (2008), $\varepsilon' = 3.16$ and $\varepsilon'' = 8 \times 10^{-3} \cdot (\nu/150\,\text{GHz})$ at a temperature of $-30°C$. In the case of nonspherical particles, the polarizability $A$ depends on the polarization direction. We will discuss this effect in the next Section. The scattering cross section given above is appropriate for Rayleigh scattering. As the ice particle diameter approaches the wavelength of the scattered radiation, Mie scattering theory provides a more accurate description (Y. Li et al. 2023). However, as described in Section 2.2.1, ice crystals this large are uncommon in the atmosphere above the South Pole.

The ice crystals scatter thermal radiation in the direction of the observer. We neglect radiation absorbed or emitted by the atmosphere and the effect from the curvature of the Earth and assume that the illumination is from a uniform plane. Thus, the scattering signal is obtained as

$$T_{RJ,sca} = \frac{1}{2}\tau_{sca}T_g, \quad (3)$$

where $T_{RJ,sca}$ is the Rayleigh–Jeans (RJ) brightness temperature of the scattered radiation, $\tau_{sca}$ is the scattering optical depth, and $T_g$ is the temperature of the ground. The factor of $1/2$ represents the fraction of the solid angle covered by the ground. The general optical depth $\tau$ is given by $\tau = nL\sigma$, where the number density $n$ and length along the line of sight $L$ will be identical in both scattering and emission, and $\sigma$ will differ as given in Equations (1) and (2). In addition to scattering, the ice crystals both absorb and emit radiation. The emission signal is obtained as

$$T_{RJ,emi} = \tau_{abs}T_p, \quad (4)$$

where $\tau_{abs}$ is the optical depth for absorption, and $T_p$ is the temperature of the ice crystals. The total Rayleigh–Jeans temperature of ice clouds is

$$T_{RJ,ice} = T_{RJ,sca} + T_{RJ,emi}. \quad (5)$$

Assuming $T_g \approx T_p$, the ratio of the scattering and emission temperature signals becomes

$$\frac{T_{RJ,sca}}{T_{RJ,emi}} = \frac{\omega^3 a^3}{9c^3}\frac{(\varepsilon'-1)^2 + \varepsilon''^2}{\varepsilon''} = \left(\frac{\nu}{\nu_0}\right)^2\left(\frac{a}{a_{eq}(\nu_0)}\right)^3. \quad (6)$$

The radius of a spherical ice crystal for which scattering and emission contribute equally, $a_{eq}(\nu_0)$, is 108, 80, and $62\,\mu m$ for $\nu_0 = 95$, 150, and 220 GHz, respectively. Scattering is dominant for particles larger than $a_{eq}$, and emission is dominant for smaller particles. Previous studies considered only scattering (L. Pietranera et al. 2007; S. Takakura et al. 2019) or emission (S. Paine 2022). In practice, however, both effects make significant contributions to the observed signal for typical ice crystal sizes.

The scattering and emission signals scale differently with frequency, and their combination results in a spectral index that depends on particle size and observing frequency. For each of the two effects, the RJ temperature $T_{RJ}(\nu) \propto \nu^\alpha$ scales as a different power of frequency. The frequency scaling of the intensity spectrum is $S(\nu) \propto \nu^{\alpha+2}$. The spectral index of the scattering signal, $\alpha_{sca} = 4$, and that of the emission signal, $\alpha_{emi} = 2$, are found from Equations (1) and (2). The signal we observe is the sum of these two components, and their relative contributions depend on the frequency.

Figure 2 shows the relative contributions and effective spectral index as a function of the particle radius $a$. The top panel shows the signals $T_{RJ,sca}$, $T_{RJ,emi}$, and their sum at 95, 150, and 220 GHz. In this plot, the ice water path (IWP), the column mass density of ice crystals, is normalized to $1\,g\,m^{-2}$ and $T_g = T_p = 240$ K. With this normalization, $T_{RJ,emi}$ is independent of the particle size, whereas $T_{RJ,sca}$ increases as $\propto a^3$ and dominates above a critical size $a_{eq}$. The bottom panel shows the effective spectral index for each pair of bands. It varies between $\alpha_{emi} = 2$ and $\alpha_{sca} = 4$ depending on the size of the ice crystals.

These calculations are for a single ice crystal size. As was described in Section 2.2.1, atmospheric ice crystals are typically observed with a distribution of sizes. The thermal emission from an ice crystal scales as its volume ($a^3$), and the Rayleigh scattering scales as volume squared ($a^6$). Therefore, the polarized signal and the resulting spectral index will be dominated by the largest ice crystals in the distribution. For this reason, care must be taken in interpreting metrics such as the relative IWP for observation sites as is presented in C.-L. Kuo (2017). The ice crystal size distribution is far more important than the total column density of ice.

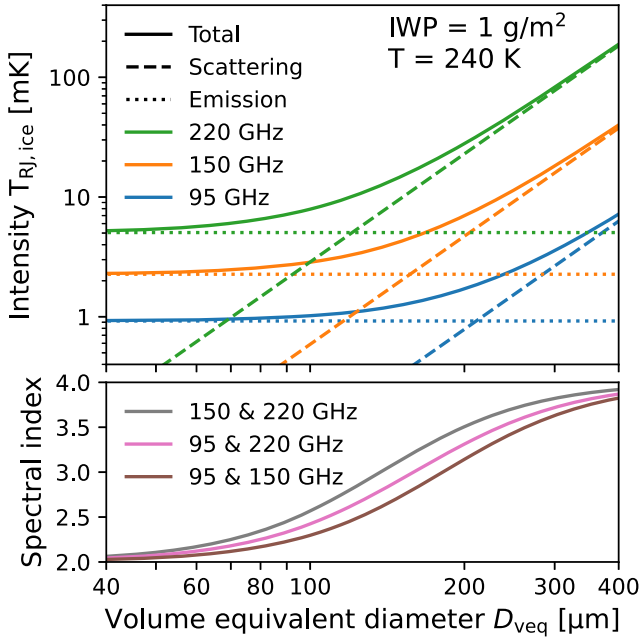

**Figure 2.** Top: temperature signal from spherical ice crystals for scattering, emission, and their sum. To get $Q$ polarization, this is multiplied by the polarization fraction. Bottom: effective spectral index of the summed temperature signal between two frequency bands.

### 3.2. Polarized Signals from Ice Crystals

Next, we consider polarization of the emitted and scattered radiation due to the nonspherical shape of the ice crystals. It is possible for the scattered radiation from spherical ice crystals to be polarized if the illumination pattern of the ice crystals has a significant quadrupole component. S. Takakura et al. (2019) considered the quadrupole illumination produced by the curvature of the Earth in the absence of atmosphere and found that it would result in a horizontal polarization fraction of <1% for spherical scatterers. However, the effect of emission from the atmosphere is important. Near the horizon, the atmosphere becomes optically thick, and the quadrupole component due to the Earth's curvature is largely canceled. The emission from the atmosphere creates an orthogonally oriented quadrupole illumination that results in vertical polarization for spherical scatterers (A. V. Troitsky et al. 2003). If the ice crystals (or water droplets) dominating the scattering in the atmosphere at the South Pole were spherical, we would expect to see vertically polarized emission with a polarization fraction of ~1%–2% in the SPT-3G observing bands. However, as we will show in this Section, for reasonable assumptions about the shape and alignment of ice crystals, horizontal polarization dominates, and we can neglect the effect of the quadrupole component of illumination.

Although ice crystals can have a variety of shapes, as was described in Section 2.2.1, here we consider the two basic and common shapes: hexagonal columns and plates. Complex shapes like rosettes could be modeled as a superposition of these basic shapes with different orientations. An important consequence of ice crystals having nonspherical shapes is their alignment. Ice crystals falling in quiescent air tend to be horizontally aligned, i.e., they face their broad side down. On the other hand, turbulence in the air can disturb this alignment. As shown by K. Gustavsson et al. (2021), only small particles with a mean radius smaller than 10 $\mu$m are randomly oriented, and larger particles are expected to be horizontally aligned. The azimuthal orientation of horizontally aligned columnar particles is assumed to be random, however, it could be possible for wind to create an azimuthal alignment.

The optical properties of nonspherical particles are represented by the complex polarizability tensor $\mathbb{A}$. For axisymmetric shapes, each component of $\mathbb{A}$ becomes

$$A_{ij} = A_\perp (\delta_{ij} - n_i n_j) + A_\parallel n_i n_j, \quad (7)$$

where $n$ is the unit vector along the symmetry axis. $A_\parallel$ and $A_\perp$ are the polarizability along the symmetric and orthogonal axes, respectively. They are calculated as

$$A_\parallel = \frac{\varepsilon - 1}{1 + (\varepsilon - 1)\Delta} \quad \text{and} \quad A_\perp = \frac{\varepsilon - 1}{1 + (\varepsilon - 1)(1 - \Delta)/2}, \quad (8)$$

where $\Delta$ is the depolarization factor along the symmetry axis. The value of $\Delta$ can be analytically calculated for spheroids: $\Delta = 0.53$ for an oblate shape with aspect ratio of 0.5, and $\Delta = 0.075$ for a prolate shape with aspect ratio of 4. Collections of particles with multiple orientations can be modeled by averaging Equation (7) over a range of values of $n$. In the case of random orientation, the polarizability becomes isotropic with $|A|^2 = (|A_\parallel|^2 + 2|A_\perp|^2)/3$ and $\text{Im}[A] = (\text{Im}[A_\parallel] + 2\text{Im}[A_\perp])/3$ for scattering and emission, respectively. In the case of horizontal alignment, the polarizability for vertical and horizontal directions are

$$|A_v|^2 = \begin{cases} |A_\perp|^2 & \text{(column)}, \\ |A_\parallel|^2 & \text{(plate)}, \end{cases} \quad (9)$$

and

$$|A_h|^2 = \begin{cases} (|A_\parallel|^2 + |A_\perp|^2)/2 & \text{(column)}, \\ |A_\perp|^2 & \text{(plate)}, \end{cases} \quad (10)$$

respectively. Here, $|A_v| \leqslant |A_h|$. Similarly, $\text{Im}[A_v]$ and $\text{Im}[A_h]$ are also calculated from $\text{Im}[A_\parallel]$ and $\text{Im}[A_\perp]$.

Due to the projection of the crystals along the line of sight to the observer, when the particles are observed at elevation $\epsilon$, the intensity of vertical polarization per particle depends on the elevation as $I_v \propto |A_v|^2 \cos^2 \epsilon + |A_h|^2 \sin^2 \epsilon$ for scattering and $I_v \propto \text{Im}[A_v]\cos^2 \epsilon + \text{Im}[A_h]\sin^2 \epsilon$ for emission. On the other hand, horizontal polarization depends on $\epsilon$ as $I_h \propto |A_h|^2$ for scattering and $I_h \propto \text{Im}[A_h]$ for emission. Thus, the polarization fraction is calculated as

$$p_\gamma \equiv \frac{Q_{\text{RJ},X}}{T_{\text{RJ},X}} = \frac{I_v - I_h}{I_v + I_h} = -\frac{(\gamma - 1)\cos^2 \epsilon}{\gamma(1 + \sin^2 \epsilon) + \cos^2 \epsilon}, \quad (11)$$

where $X$ denotes scattering or emission, $Q_{\text{RJ},X}(<0)$ is the Stokes parameter for horizontal linear polarization, and

$$\gamma = \begin{cases} |A_h|^2/|A_v|^2 & \text{(scattering)}, \\ \text{Im}[A_h]/\text{Im}[A_v] & \text{(emission)}. \end{cases} \quad (12)$$

Here, $\gamma$ is similar for scattering and emission because $\varepsilon'' \ll 1$, and is only dependent on the crystal geometry: $\gamma = 2.0$ for both the oblate (plate) shape with an aspect ratio of 0.5 and for the prolate (column) shape with an aspect ratio of 4.

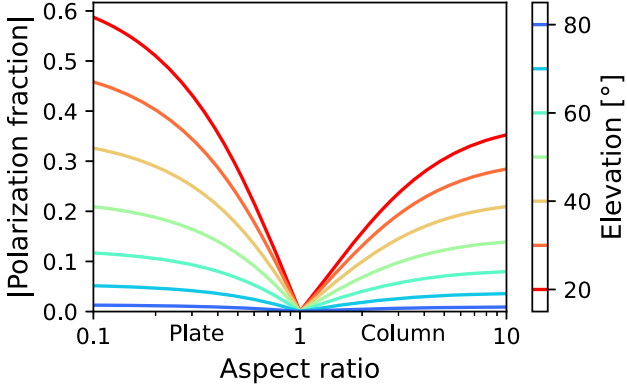

**Figure 3.** Polarization fraction of the signal from ice crystals with different shapes at different observing elevations.

The polarization fractions for scattering and emission with $\gamma = 2.0$ are similar, so the spectral index of the polarized signal from ice will be very close to that of the unpolarized signal given in Figure 2. Therefore, the spectral index we measure for the polarized anisotropy power has the potential to constrain the size of the atmospheric ice crystals responsible for the dominant contribution to the signal.

Figure 3 shows the polarization fraction calculated from Equation (11). The polarization fraction is minimized for random orientation, small aspect ratio, and observations near the zenith. For observations at the zenith, horizontally oriented ice crystals present an azimuthally symmetric distribution of ice with a vanishing polarization fraction.

## 4. Atmospheric Power Modeling

In previous studies of atmospheric temperature anisotropy by O. P. Lay & N. W. Halverson (2000), R. S. Bussmann et al. (2005), J. Sayers et al. (2010), and J. Errard et al. (2015), the 3D temperature anisotropy power was assumed to scale with spatial wavenumber $k$ as

$$P(k) \propto k^{-\beta}. \quad (13)$$

In the Kolmogorov theory of turbulence, the exponent of this scaling is $\beta = 11/3$; however, in the following discussion, we will leave $\beta$ as a free parameter unless stated otherwise. In this work, we adopt a simple model where the 3D anisotropic distribution can be integrated along the line of sight to model the emission as a 2D screen. We will see that this model provides an excellent explanation of the observed spatial scaling of power.

We treat the emission as being optically thin and coming from a layer of thickness $\Delta h$ at a height $h$. In this limit, the 2D angular power spectrum can be expressed as

$$P(\alpha_x, \alpha_y) = BT_\nu^2 (\sin \epsilon)^{1-\beta} (\alpha_x^2 + \alpha_y^2)^{-\beta/2}, \quad (14)$$

where $\alpha_x = k_x h / \sin \epsilon$ and $\alpha_y = k_y h / \sin \epsilon$ are angular wavenumbers and $BT_\nu^2$ is the amplitude of the TT angular power spectrum normalized to observations at the zenith. In previous work, such as R. S. Bussmann et al. (2005), this quantity is written as $B_\nu^2$. Here, we adopt $BT_\nu^2$ to differentiate between the amplitude of temperature and polarization fluctuations. $BT_\nu^2$

depends only on the properties of the atmosphere and is related to the RJ temperature of the anisotropic emission/scattering by water vapor/ice per unit thickness of atmosphere as a function of frequency $\kappa_\nu$, height $h$, and thickness $\Delta h$ of the atmospheric layer,

$$BT_\nu^2 \propto \kappa_\nu^2 h^{\beta-2} \Delta h, \quad (15)$$

and has units of $mK^2 \, rad^{2-\beta}$. The geometric factor of $(\sin \epsilon)^{1-\beta}$ accounts for the dependence on elevation of the path length through the atmospheric layer and the physical scale of fluctuations being probed.

In the case of Kolmogorov turbulence with $\beta = 11/3$, the 2D angular power spectrum becomes

$$P(\alpha_x, \alpha_y) = BT_\nu^2 (\sin \epsilon)^{-8/3} (\alpha_x^2 + \alpha_y^2)^{-11/6} \quad (16)$$

with

$$BT_\nu^2 \propto \kappa_\nu^2 h^{5/3} \Delta h. \quad (17)$$

In this work, we measure the 1D angular power spectra from the cross correlations of Stokes $T$, $Q$, and $U$ time-ordered data (TOD). We assume that the scan speed is much faster than the wind speed and that sky can be assumed to be stationary or "frozen." If the sky has a 2D angular power spectrum of $P(\alpha_x, \alpha_y)$, the power spectrum measured by an experiment scanning in an infinitesimally narrow strip in the $x$-direction is

$$\begin{aligned} P(\alpha_x) &= \int_{-\infty}^{\infty} P(\alpha_x, \alpha_y) d\alpha_y \\ &= BT_\nu^2 (\sin \epsilon)^{1-\beta} \int_{-\infty}^{\infty} (\alpha_x^2 + \alpha_y^2)^{-\beta/2} d\alpha_y \\ &= BT_\nu^2 (\sin \epsilon)^{1-\beta} \frac{\sqrt{\pi} \Gamma(\frac{\beta}{2} - \frac{1}{2})}{\Gamma(\frac{\beta}{2})} \alpha_x^{1-\beta}. \end{aligned} \quad (18)$$

We have ignored the convolution with the ~1′ FWHM beam, as the measured signal is dominated by the much larger power on degree angular scales. In the case of a Kolmogorov fluctuation spectrum with $\beta = 11/3$, we have

$$P(\alpha_x) = BT_\nu^2 (\sin \epsilon)^{-8/3} \frac{\sqrt{\pi} \Gamma(\frac{4}{3})}{\Gamma(\frac{11}{6})} \alpha_x^{-\frac{8}{3}}. \quad (19)$$

With Equation (19), we can use the measured SPT-3G 1D cross spectra to constrain the instantaneous amplitude and spatial scaling of the atmospheric temperature anisotropy power.

### 4.1. Extending to Polarized Signal

The polarization fraction of the signal due to ice, $p_\gamma$, depends on the observation elevation. It is given by Equation (11) and shown in Figure 3. In this work, we will characterize the properties of the atmosphere independent of the observing elevation and provide a prescription for simulating the polarized sky at any elevation.

We parameterize the elevation-independent polarized power amplitude as a function of frequency by $BQ_\nu^2$. This quantity has the same physical dependencies as $BT_\nu^2$ in Equation (17) with the exception that only emission/scattering from ice (not water vapor) contributes to the signal. The value of $BQ_\nu^2$ depends on both the column depth of ice crystals and very sensitively on the distribution of their sizes. The measured $Q$ power spectrum will include an additional factor of $p_\gamma^2(\epsilon)$ to account for the polarization fraction of the signal from ice as a function of ice crystal shape ($\gamma$) and elevation of observation ($\epsilon$). In this case, we have

$$P(\alpha_x, \alpha_y) = BQ_\nu^2 p_\gamma^2(\epsilon)(\sin \epsilon)^{1-\beta}(\alpha_x^2 + \alpha_y^2)^{-\beta/2}$$
$$= BQ_\nu^2 f(\epsilon)(\alpha_x^2 + \alpha_y^2)^{-\beta/2}, \quad (20)$$

where the factor

$$f(\epsilon) = p_\gamma^2(\epsilon)(\sin \epsilon)^{1-\beta} = \frac{(\gamma - 1)^2 \cos^4 \epsilon (\sin \epsilon)^{1-\beta}}{[\gamma(1 + \sin^2 \epsilon) + \cos^2 \epsilon]^2}, \quad (21)$$

encodes the complete dependence of the QQ power spectrum on elevation. Similar to the expression for the temperature fluctuation power, $BQ_\nu^2$ represents the elevation independent amplitude of the QQ power and has units of $mK^2 \, rad^{2-\beta}$. The dependence on gamma is relatively soft, and we leave $\gamma = 2$ fixed. With the reasonable assumptions of $\gamma = 2$ and $\beta = 11/3$, the observed QQ power scales with elevation as

$$f(\epsilon) = \frac{\cos^4 \epsilon (\sin \epsilon)^{-8/3}}{(3 + \sin^2 \epsilon)^2}. \quad (22)$$

The observed polarized power declines steeply with increasing elevation and vanishes for observations at the zenith. QQ power near the bottom of the SPT-3G 1500 $deg^2$ survey at elevation $\epsilon = 44°.75$ is predicted to be $\sim 28.4$ times higher than that near the top of the survey at $\epsilon = 67°.25$.

In the case of a Kolmogorov fluctuation spectrum with $\beta = 11/3$, we have

$$P(\alpha_x) = BQ_\nu^2 f(\epsilon) \frac{\sqrt{\pi} \, \Gamma(\frac{4}{3})}{\Gamma(\frac{11}{6})} \alpha_x^{-\frac{8}{3}}. \quad (23)$$

Equation (23) is only formally correct in the case of a $Q$ signal arising from polarized atmosphere; however, we define and analyze $BU_\nu^2$ identically for comparison purposes.

## 5. Instrument and Data Set

The data in this paper were taken with the SPT-3G instrument on the 10 m diameter SPT between the end of March and the beginning of October in the 2019, 2020, 2021, and 2022 Austral winter observing seasons. The SPT is located $\sim 1$ km from the geographic South Pole, and observing elevation, $\epsilon$, is related to source decl., $\delta$, as $\epsilon = -\delta$.

SPT-3G is the third-generation survey instrument on the SPT and was deployed in 2017. The SPT-3G camera consists of $\sim 16,000$ polarization sensitive detectors, distributed over ten 150 mm silicon wafers. Each wafer contains 269 pixels, each of which has a dual-polarized broadband sinuous antenna. These antennas feed filter banks that separate the signals in each polarization into bands centered at 95, 150, and 220 GHz. Transition edge sensor bolometers measure the incoming power in each of six combinations of frequency band and polarization. The pixels on the detector wafers are evenly distributed between two polarization angles corresponding to relative orientations of 0° or 45°. Each of these is paired with a mirror image pixel, which, when averaged, cancels any polarization rotation with changing frequency in the antenna. Each of the 10 wafers has one of six relative orientations, resulting in a distribution of detector orientation angles spaced 15° apart.

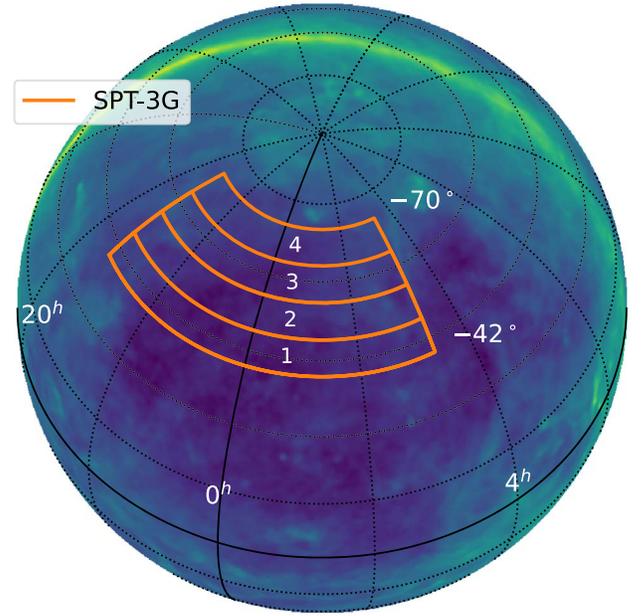

**Figure 4.** SPT-3G subfields overlaid on a thermal dust map from Planck (Planck Collaboration et al. 2016). Subfields 1, 2, 3, and 4 correspond to fields centered at decl. $\delta = -44°.75$, $-52°.25$, $-59°.75$, and $-67°.25$, respectively. Each subfield is 7°.5 tall and 100° wide.

The observations used in this work cover a $\sim 1500 \, deg^2$ region extending from $-42°$ to $-70°$ decl. and from $20^h 40^m 0^s$ to $3^h 20^m 0^s$ R.A. This $\sim 1500 \, deg^2$ survey is divided in elevation into four 7°.5-tall subfields (shown in Figure 4) centered at decl. $\delta = -44°.75$, $-52°.25$, $-59°.75$, and $-67°.25$ and covering the complete R.A. range. Each subfield is observed in a raster-scan pattern, making sweeps in azimuth of constant elevation. Each sweep, referred to as a "scan," takes approximately 100 s and covers the full azimuth range of 100°. The telescope steps in elevation after each scan pair (one left-going, one right-going). The duration of each subfield observation is approximately 2 hr. During each observing day, defined by the combined fridge hold time and cycle time, two subfields are observed three times each. More information on the SPT-3G instrument and survey can be found in J. A. Sobrin et al. (2022).

### 5.1. Data Processing

The bolometer TOD are divided by scan and referred to as "timestreams." For each scan, we decompose the timestreams into the Stokes parameters $T$, $Q$, and $U$ for each wafer and frequency band using Equation (24).

$$X = \begin{pmatrix} \sum_i 1 & \sum_i \cos(2\psi_i) & \sum_i \sin(2\psi_i) \\ \sum_i \cos(2\psi_i) & \sum_i \cos^2(2\psi_i) & \sum_i \cos(2\psi_i) \sin(2\psi_i) \\ \sum_i \sin(2\psi_i) & \sum_i \sin(2\psi_i) \cos(2\psi_i) & \sum_i \sin^2(2\psi_i) \end{pmatrix}$$

$$Y = \begin{pmatrix} \sum_i y_i \\ \sum_i y_i \cos(2\psi_i) \\ \sum_i y_i \sin(2\psi_i) \end{pmatrix}$$

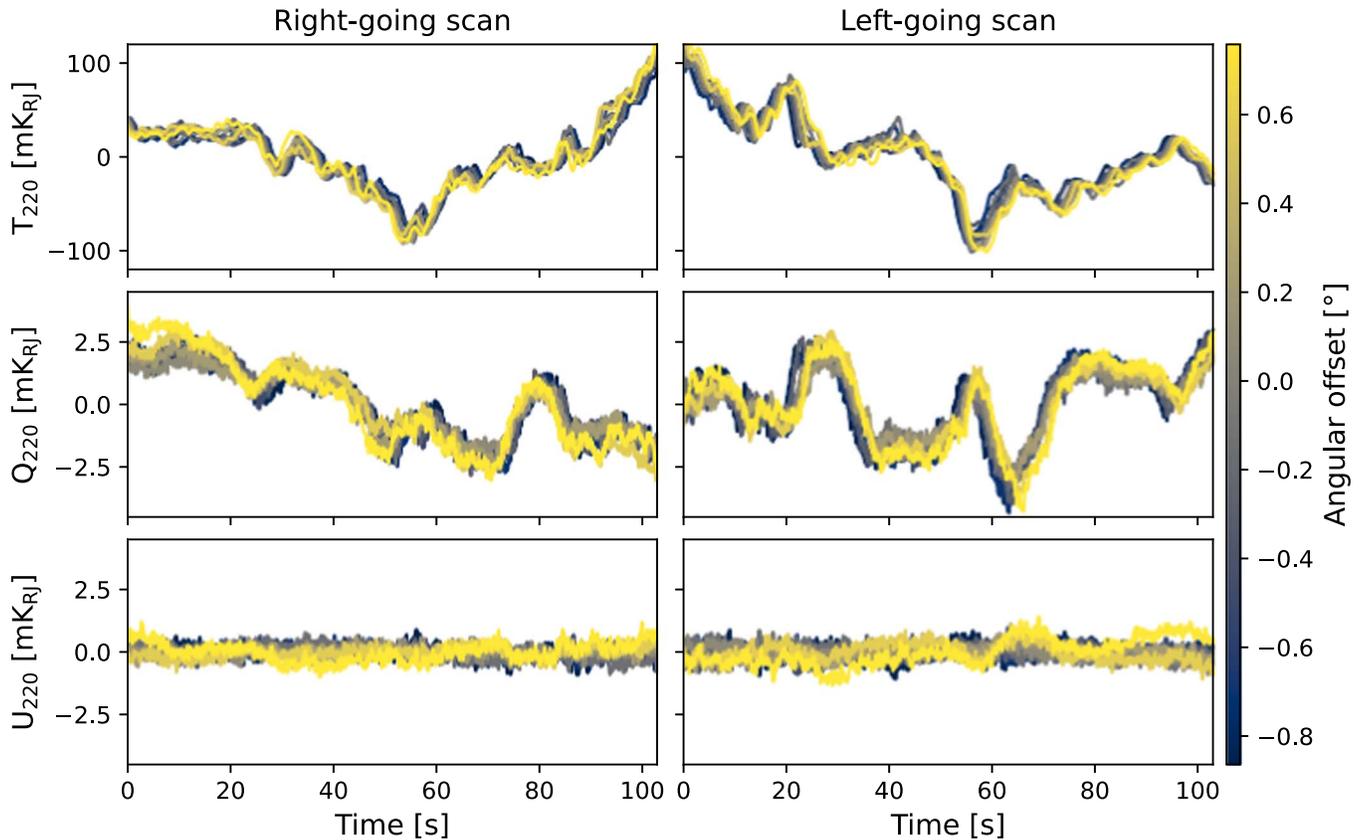

**Figure 5.** Example timestreams in the presence of a large $Q$ polarized signal. The left and right panels show consecutive right-going and left-going scans, and the top, middle, and bottom panels show the $T$, $Q$, and $U$ signals, respectively. Each line shows averaged timestreams from each of the nine detector wafers, whose color indicates the horizontal angular offset of the wafer from the telescope boresight. These timestreams have been mean subtracted and are insensitive to any DC component of the temperature or polarization. The $Q$ signal is dominated by low temporal frequencies and is much larger than $U$, which is expected to be zero for horizontal polarization. The larger $T$ signal is effectively suppressed by detector differencing. For both $T$ and $Q$, the timestreams of the nine independent detector wafers are highly correlated, but are temporally lagged due to a combination of wind speed and, to a greater extent, the motion of the telescope.

$$\begin{pmatrix} T \\ Q \\ U \end{pmatrix} = X^{-1} Y \qquad (24)$$

Here, $y_i$ is the detector TOD, and $\psi_i$ is polarization angle of the detector. We only use detectors that have a functional orthogonal polarization pair in the same spatial pixel.

In this analysis, we measure the atmospheric fluctuations over a broad range of angular scales with preference to large scales where the signal is larger. To achieve this goal, the data are only lightly filtered to preserve signal on large angular scales. The bolometer timestreams have an offset and linear drift removed over each of the $100°$ azimuth scans. Filtering choices are identical for the $T$, $Q$, and $U$ timestreams. The detector gains are matched by adjusting them to have a uniform response to the temperature signal produced by moving the telescope in elevation. Typically, this constrains the relative gain of each detector to an accuracy of $\sim 1\%$. In the limit where these uncertainties are the result of random uncorrelated errors, we expect the leakage of temperature to polarization to be $1\%/\sqrt{N_{\rm det}} \sim 0.01\%$.

The analysis presented here is based on cross spectra between independent detector wafers in the SPT-3G focal plane. The assumption that the atmospheric signal is completely correlated between detector wafers will begin to break down as we probe scales corresponding to the angular separation of wafers. The SPT-3G focal plane subtends a solid angle of roughly $2°\!.1 \times 1°\!.7$ on the sky, so cross spectra between detectors at the edges of the most widely spaced wafers will begin to experience decorrelation on scales of $\Delta\theta < 2°\!.1$ or $\ell > 87$. However, for the mean cross spectra from all wafer pairs, this decorrelation will not become significant until $\ell \gtrsim 100$.

Figure 5 shows example timestreams from an observation with significant signal in $Q$ polarization. The signals in $T$ and $Q$ are highly correlated over the focal plane but with small delays depending on the scan direction and the pointing offset of each wafer, which strongly indicates that the signals are on the sky. In the $U$ signal, on the other hand, there is no clear correlated large angular scale structure.

We want to measure the common atmospheric signal between wafers and minimize the noise bias contributed by uncorrelated noise between wafers and frequency bands. This is accomplished by computing cross power spectra between the timestreams from the different detector wafers. We include a total of nine SPT-3G wafers in this analysis, omitting one of the 10 wafers because of poor low-frequency noise properties. For each azimuth scan, we calculate cross spectra between all wafers for each of the Stokes parameters and three observing frequency bands. This analysis is conducted in Rayleigh–Jeans temperature units. A single cross spectrum (one scan, one wafer pair) can be described by

$$\Gamma^{ij} = a \quad b_\nu^2 \quad F\{w_{\rm H} \quad x_i\} F^*\{w_{\rm H} \quad x_j\}, \qquad (25)$$

where $F$ denotes the Fourier transform, $x_i$ and $x_j$ denote detector-averaged timestreams from two wafers ($i$, $j$), $b_\nu$ is a band-dependent correction factor applied to convert the power into Rayleigh–Jeans units, and $w_H$ denotes the Hamming window function. The normalization factor $a$ is given by

$$a = \frac{2}{N \sum w_H^2} \quad (26)$$

where $N$ is the length of vectors $x_i$ and $x_j$.

For single-band cross spectra, we use the by-wafer timestreams to calculate the cross spectrum for every possible wafer pair (36 total). These 36 cross spectra are then averaged to create a single cross spectrum (one each for TT, QQ, UU) for each scan in each observation. Each scan cross spectrum $\ell$ bin is given by

$$\Gamma_\ell = \sum_{\{i,j\}} \Gamma_\ell^{ij} / N_p \quad (27)$$

where $N_p$ is the number of unique cross spectra.

Similarly, we calculate the cross-band spectra by taking the cross spectrum of every possible wafer pair between the two relevant bands (72 total). For one cross spectrum (two wafers from two frequency bands), this can be written as

$$\Gamma_{\nu_1 \nu_2}^{ij} = a \ \ b_{\nu_1} b_{\nu_2} F\{w_H \ \ x_i^{\nu_1}\} F^*\{w_H \ \ x_j^{\nu_2}\}, \quad (28)$$

where $x_i^{\nu_1}$ and $x_j^{\nu_2}$ denote detector-averaged timestreams from two wafers ($i$, $j$) and two frequency bands ($\nu_1$, $\nu_2$), and $b_{\nu_1}$ and $b_{\nu_2}$ are the band-dependent correction factors. As in the single-band case, we average the resultant cross spectra to create a single scan cross spectrum for each frequency band pair.

The uncertainty on each cross spectrum bin is estimated, in the high signal-to-noise limit, to be

$$\sigma_\ell = \sqrt{\frac{\sum_k^{N_p}(\Gamma_{\ell,k} - \overline{\Gamma}_\ell)^2}{2 N_w (N_p - 1)}} = \frac{\sigma_{\Gamma_\ell}}{\sqrt{2 N_w}}, \quad (29)$$

where $N_w$ is the number of wafers, $N_p$ is the number of unique cross spectra, $\Gamma_{\ell k}$ is the cross spectrum power for bin $\ell$, and $\sigma_{\Gamma_\ell}$ is the measured standard deviation of the cross spectrum values in that $\ell$ bin about the mean. Finally, all of the scans in an observation are averaged further to create a single cross spectrum for each observation. We have

$$XX_\ell = \sum_N \Gamma_{\ell, N} / N, \quad (30)$$

where $N$ is the number of scans in an observation, and $XX$ represents the QQ, UU, or TT observation cross power spectrum. Per-observation error is estimated from the mean variance of the scans divided by the number of scans.

## 6. Wind Speed

### 6.1. Frozen Sky Approximation

We model the anisotropic atmospheric polarization as originating from a layer of emission at a fixed height above the ground. In principle, this emitting layer can evolve in time through the redistribution of water vapor and ice crystals. Time evolution of the atmospheric emission structure at the Chajnantor site has been studied by T. W. Morris et al. (2022) and J. Errard et al. (2015), but is expected to be a very subdominant effect for the observations studied here. As was demonstrated in R. S. Bussmann et al. (2005), for the atmospheric conditions observed at the South Pole, it is an excellent approximation to model the atmospheric emission as originating from a screen of emission that is moved parallel to the ground by wind. R. S. Bussmann et al. (2005) also showed that the wind above the South Pole was typically constant in speed and direction over the course of a several-hour observation. The spatial scales of the atmospheric emission probed by a telescope scan depend on both the scan and wind angular velocities. In the limit where the angular scan speed is much greater than the angular wind speed, the wind speed can be neglected, and the anisotropic atmospheric emission can be treated as stationary. However, even low wind speeds will result in the screen of emission moving by several degrees in the time that the array completes a full 100° azimuth scan. Therefore, we can treat each scan as a new and effectively independent realization of the atmosphere.

Our goal is to measure the spatial fluctuation power of the atmosphere. In the limit where the angular scan speed of the telescope is much greater than the angular wind speed with which the atmospheric fluctuations are moved by wind, the measured angular power spectrum will reflect the true spatial power spectrum. We refer to this set of conditions as the "frozen sky" approximation. We will show in Section 6.2 that, for observations at the South Pole with the SPT-3G scanning strategy, this condition is approached but not rigorously satisfied. When the wind speed approaches or exceeds the scan speed, the angular scales of the azimuth scan will not simply correspond to spatial scales of the atmosphere. Due to the steep spatial scaling of the atmospheric power, this will result in larger observed power for scans oriented against the wind direction and a bias toward higher mean power computed from the average of all scans. We will argue that this bias is modest and well understood, and that we can interpret our measured power amplitudes as upper limits to the true atmospheric power.

### 6.2. Estimation of the Angular Wind Speed

Here, we estimate the wind speed using the temperature and polarization data from SPT-3G, and show that the wind speed is typically slower than the scan speed, and thus the "frozen sky" approximation is reasonable, although not always rigorously satisfied.

We use the $T$ or $Q$ timestream from each of the nine wafers of the SPT-3G focal plane as described in Section 5.1. The detector timestreams result from the telescope scanning over the pattern of atmospheric emission being moved by wind. We describe our method in terms of $T$, but the procedure is identical for $Q$. The spatial gradient on the sky and the temporal derivative in the detector timestreams are related as

$$\frac{\partial T}{\partial t} = (v_{Az} \cos \epsilon + w_n \sin Az - w_e \cos Az) \nabla_x \ T$$
$$+ (w_n \cos Az + w_e \sin Az) \sin \epsilon \ \nabla_y \ T, \quad (31)$$

where Az is the azimuthal angle of the telescope, $v_{Az}$ is the angular speed of the scan in azimuth, $w_n$ and $w_e$ are the angular speed of the wind toward north (Az = 0°) and east, respectively, and $x$ and $y$ are horizontal and vertical angular offsets from the telescope boresight. We first obtain focal-plane-averaged timestreams from wafer-averaged timestreams

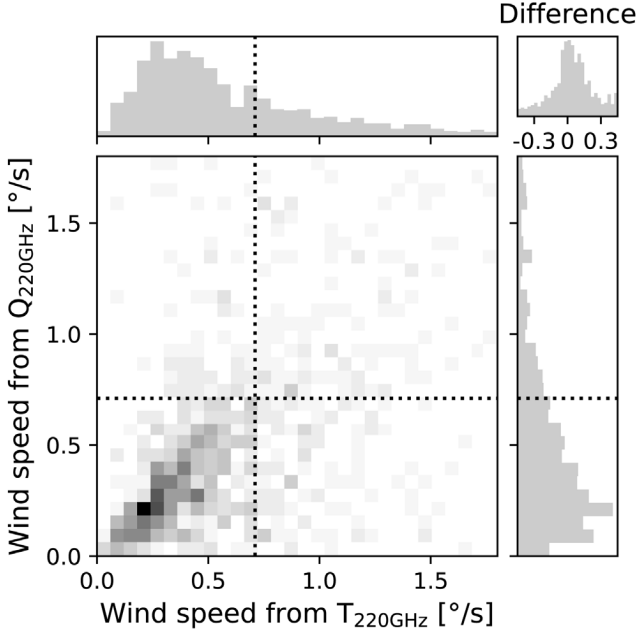

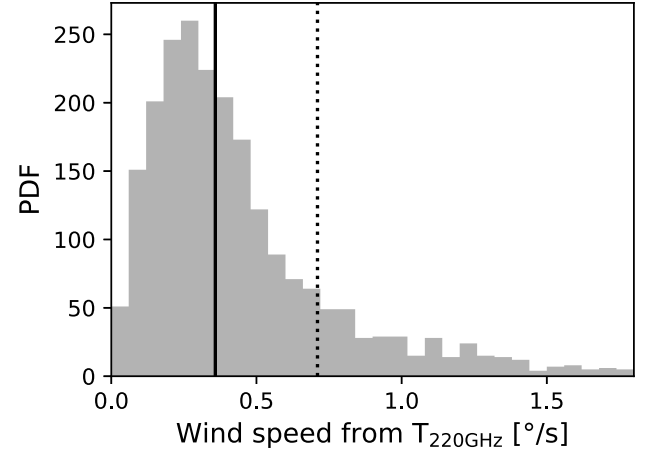

**Figure 7.** Histogram of wind speeds measured from $T$ (intensity) fluctuations for every observation. The dotted line shows the apparent angular speed of the scan for the $\delta = -44\overset{\circ}{.}75$ subfield. The median (solid line) and modal wind speeds are $0\overset{\circ}{.}36\,\mathrm{s}^{-1}$ and $0\overset{\circ}{.}26\,\mathrm{s}^{-1}$, respectively. The majority of observations have wind speeds below the apparent scan speed, validating the "frozen sky" approximation.

**Figure 6.** Comparison between the wind speeds estimated from the 220 GHz $T$ (intensity) and $Q$ (polarization) timestreams for a sample of high-$Q$ observations. The center panel shows the 2D histogram of observations with significant polarization signals, and the top and right panels show its projected histograms. The upper right panel shows the histogram of the difference. The dotted lines show the apparent angular speed of the scan for the $\delta = -44\overset{\circ}{.}75$ subfield.

$T_i$ by computing

$$\begin{pmatrix} T \\ \nabla_x T \\ \nabla_y T \end{pmatrix} = [P^T N^{-1} P]^{-1} P^T N^{-1} T_i, \quad (32)$$

where $P = (1 \quad \Delta x_i \quad \Delta y_i)$ is the pointing matrix, and $N$ is the noise covariance matrix. The time derivative $\frac{\partial T}{\partial t}$ is then computed from the focal-plane-averaged timestream. Finally, we estimate $w_n$ and $w_e$ from the correlation among $\frac{\partial T}{\partial t}$, $\nabla_x T$, and $\nabla_y T$ using Equation (31). We perform this estimation for individual right-going or left-going scans, and then take an average among scans for each 2 hr observation.

Note that this method only returns reasonable velocities when $\frac{\partial T}{\partial t}$, $\nabla_y T$, and $\nabla_x T$ are all signal-dominated. This condition is always satisfied for $T$. For $Q$, on the other hand, we can only measure the wind speed in scans with significant $Q$ signals like those seen in Figure 5. However, we are interested in the wind velocity for the emission dominating $T$ and $Q$ in every observation. To determine this, we first show that the wind speeds measured by the $T$ and $Q$ timestreams in an observation are highly correlated. This strongly suggests that the atmospheric signals dominating both $T$ and $Q$ timestreams come from a common layer in the atmosphere and will typically move with similar angular velocities. We first select observations with large 220 GHz $Q$ power ($>1\,\mathrm{mK}\sqrt{s}$) and then compute the average wind speed for all scans in each observation. Figure 6 shows a comparison of angular wind speed, $\sqrt{w_n^2 + w_e^2}$, estimated from the $T$ and $Q$ timestreams at 220 GHz. The median wind angular speeds from $T$ and $Q$ are found to be $0\overset{\circ}{.}52\,\mathrm{s}^{-1}$ and $0\overset{\circ}{.}46\,\mathrm{s}^{-1}$, respectively, with the measured values being highly correlated.

We anticipate that large $Q$ signals could be correlated with high wind speed. Therefore, rather than using the median wind speed determined from the high-$Q$ subsample, we use the median wind speed measured by $T$ to characterize the typical conditions. In Figure 7, we show a histogram of the wind speeds measured from the $T$ timestreams for every observation. As anticipated, the $T$ wind speeds are lower than those from the high-$Q$ observation sample and have a median wind speed of $0\overset{\circ}{.}36\,\mathrm{s}^{-1}$.

All power spectrum amplitude measurements come from observations of the lowest ($\delta = -44\overset{\circ}{.}75$) subfield where the scan speed of $v_{\mathrm{Az}} \cos(44.75) = 0\overset{\circ}{.}71\,\mathrm{s}^{-1}$ is typically faster than the wind velocity, and the frozen sky approximation is reasonable. The effective angular speed with which the telescope scans over atmospheric structure on the sky is given by

$$v_{\mathrm{eff}} = [(v_{\mathrm{Az}} \cos \epsilon + \omega_n \sin \mathrm{Az} - \omega_e \cos \mathrm{Az})^2 \\ + (\omega_n \cos \mathrm{Az} + \omega_e \sin \mathrm{Az})^2 \sin^2 \epsilon]^{1/2}. \quad (33)$$

If $v_{\mathrm{eff}} > v_{\mathrm{Az}} \cos \epsilon$, the measured power will correspond to larger physical scales than implied by the telescope scan speed. If the atmospheric power is described by the spatial scaling $P(\alpha) \propto \alpha^{-8/3}$, then we can calculate the amount by which the power is overestimated from ignoring the motion due to wind. In a single scan, the ratio of observed power to true sky power will be scaled by a factor $g = (v_{\mathrm{eff}}/v)^{8/3}$. Since our observations consist of both left ($+v_{\mathrm{Az}}$) and right ($-v_{\mathrm{Az}}$) scans, the average for a left/right scan pair is

$$\bar{g} = \frac{1}{\Delta \mathrm{Az}} \times \int_{-\Delta \mathrm{Az}/2}^{\Delta \mathrm{Az}/2} \frac{v_{\mathrm{eff}}(+v_{\mathrm{Az}})^{8/3} + v_{\mathrm{eff}}(-v_{\mathrm{Az}})^{8/3}}{2(v_{\mathrm{Az}} \cos \epsilon)^{8/3}} d\mathrm{Az}. \quad (34)$$

Assuming the measured median wind speed of $0\overset{\circ}{.}36\,\mathrm{s}^{-1}$, we estimate that the measured QQ and TT power amplitudes will overestimate the true power by a factor of $\bar{g} = 1.48$ for wind blowing parallel to the scan direction at the center of the scan and $\bar{g} = 1.24$ for wind blowing perpendicular to the scan direction at the center of the scan. Assuming the median wind speed and averaging over wind direction, we expect the reported TT and QQ power amplitudes to exceed the true power on the sky by ∼37%. We do not correct for this bias and present the measured power amplitudes with the caveat that they are to be treated as upper limits to the true sky power. For observations of the higher elevation fields, the slower scan speeds will result in significant overestimates of the true sky power.

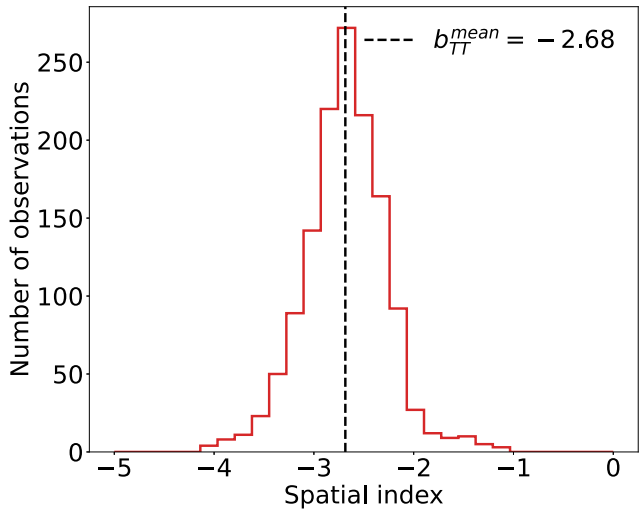

Figure 8. Histogram of spatial scaling indices of the TT signal, determined from the averaged 220 GHz cross spectra from 2 hr observations of the $\delta = -44\overset{\circ}{.}75$ subfield.

## 7. Characterization of Temperature Anisotropy

In R. S. Bussmann et al. (2005), the authors used the ACBAR experiment to measure histograms of $BT_\nu^2$ for frequency bands centered at $\nu = 150$, 220, and 278 GHz for Austral winter observations at the South Pole. These histograms have been used for detailed modeling of atmospheric temperature anisotropy at the South Pole. In this Section, we analyze our TT data identically to QQ (presented in Section 8) and compare our results with those of R. S. Bussmann et al. (2005) as a robust test of our analysis method. We first fit the measured 220 GHz wafer TT cross spectra and show that the spatial scaling of the atmospheric temperature anisotropy is consistent with the predicted Kolmogorov power-law scaling. We then measure and report the amplitude of the TT anisotropy power in each of the three bands for every observation. For the 150 and 220 GHz bands, we find our amplitude histograms are generally consistent with those of R. S. Bussmann et al. (2005) despite the differences in analysis method and observation period. Additionally, we provide results for the 95 GHz band that were not available with the ACBAR experiment.

### 7.1. Spatial Scaling of Temperature Anisotropy

Temperature anisotropy is expected to follow a Kolmogorov scaling with angular scale. Equations (19) and (23) predict that the measured 1D cross spectra will scale as

$$P(\alpha) \propto \alpha^{1-\beta}. \tag{35}$$

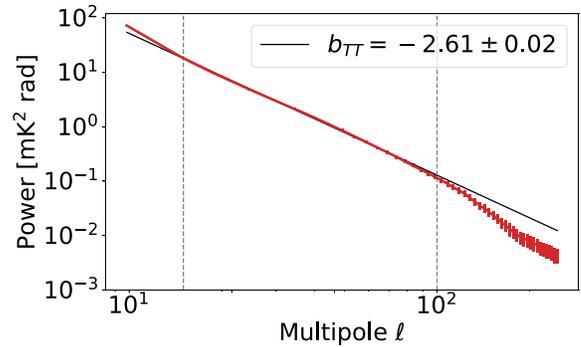

Figure 9. Spatial scaling of the 220 GHz cross spectra from the average of all observations of the $\delta = -44\overset{\circ}{.}75$ subfield. The vertical gray dotted lines denote the fit range. The spatial scaling is remarkably well fit by a power law consistent with Kolmogorov turbulence. The deviation from the power-law scaling for $\ell > 100$ is due to decorrelation caused by the finite separation of wafers in the SPT-3G focal plane.

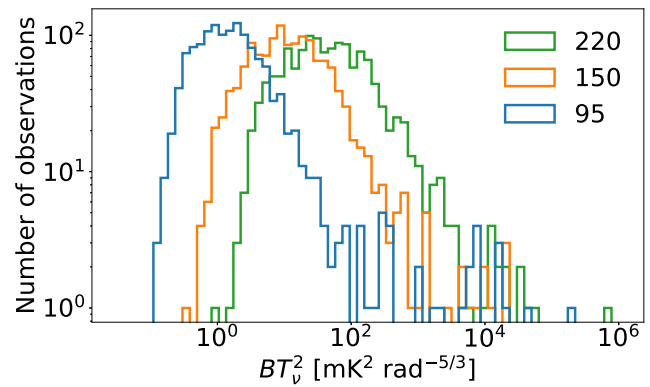

Figure 10. Histograms of temperature power amplitude for each observation in each of the three frequency bands.

We define $b = 1 - \beta$, where $b = -8/3$ for Kolmogorov turbulence.

The average 220 GHz cross spectra computed from 2 hr observations of the $\delta = -44\overset{\circ}{.}75$ subfield are fit with a power law using nonlinear least-squares over a range of $15 < \ell < 100$. We express the range of the fit in terms of spherical harmonic multipole order, $\ell = 2\pi\alpha$, due to its widespread use in CMB analysis. This range is chosen to avoid the limits set by the angular extent of the telescope azimuth scan and the size of the SPT-3G focal plane.

In Figure 8, we present a histogram of the resultant spatial scaling values for all observations. The spatial scaling for each observation is well fit by a power law, and the standard deviation of the individual measurements from the mean is $\sigma_b = 0.42$. The mean value of the scaling, $b_{TT}^{mean} = -2.68 \pm 0.01$, is in excellent agreement with the predicted Kolmogorov scaling described in Section 4.

In order to demonstrate how well the observed power is approximated by a Kolmogorov spectrum, we show the average TT power spectrum in Figure 9. The spectrum shown is an unweighted average of the bottom 80% of TT observation spectra. The top 20% of observations are discarded to avoid a few high-amplitude observations dominating the result. Error bars for each $\ell$ bin in the summed power spectrum are calculated by combining the errors for that bin from each observation in quadrature and dividing by the number of observations. The spectrum is fit, as above, over the range $15 < \ell < 100$. The errors on each $\ell$ bin are uniformly scaled so that the reduced $\chi^2 = 1$ for this fit to the power-law model. The average power spectrum spatial scaling is fit by a power law with $b_{TT} = -2.61 \pm 0.02$.

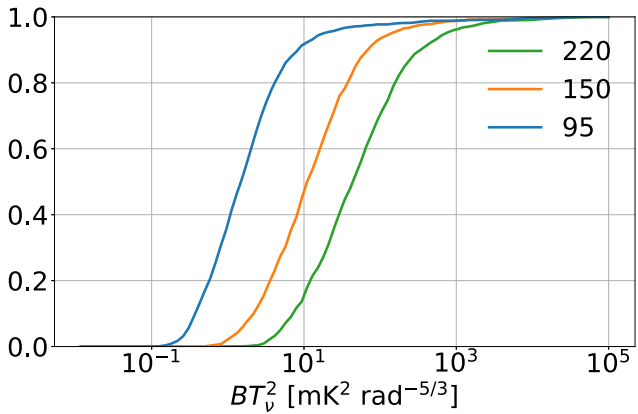

**Figure 11.** Cumulative distribution functions (CDFs) of the temperature power amplitude from each 2 hr observation for each of the three frequency bands.

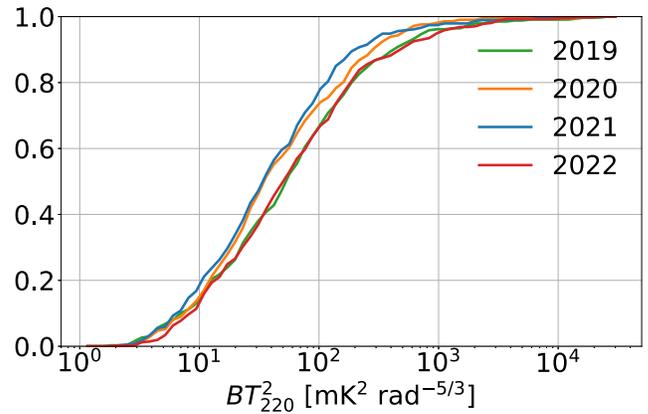

**Figure 12.** CDFs of the temperature power amplitude at 220 GHz from each observation broken up by year of observation.

### 7.2. Temperature Anisotropy Amplitude Distribution

We use Equation 19 to fit for the amplitude of the cross spectrum power in the three frequency bands for every 2 hr observation of the $\delta = -44^\circ\!.75$ subfield. We express the TT power amplitude as the quantity $BT_\nu^2$, which characterizes the amplitude of the atmospheric anisotropic emission and is independent of both observing elevation and spatial scale. Histograms of TT power amplitude for each observation in the 95, 150, and 220 GHz bands are shown in Figure 10. It is worth noting that significant temperature anisotropy power is detected in every observation regardless of the weather. The sensitivity of SPT-3G to large-scale CMB temperature anisotropy is limited by atmospheric noise in all three observing bands. This robust detection of atmospheric signal makes it clear that the temperature signal must be highly suppressed if we do not want our measurements of CMB polarization to be limited by temperature to polarization leakage from the atmosphere. These histograms are used to create cumulative distribution functions (CDFs) for the TT power measured in each of the three frequency bands for every observation, shown in Figure 11. Percentile values from these CDFs are given in Table 1, and a more complete sampling of the CDF can be found in Appendix A. Despite differences in analysis methods and concerns about the impact of wind speed, these measurements of $BT_\nu^2$ are shown to be generally consistent with the results of R. S. Bussmann et al. (2005). In Figure 12, we show the CDFs of $BT_{220}^2$ for each of the SPT-3G observing years and find that the atmospheric conditions are comparable for all 4 yr.

### 8. Characterization of Atmospheric Polarization Anisotropy with SPT-3G

Here, we characterize the properties of the atmospheric polarization anisotropy at the South Pole. In Section 8.1, we qualitatively describe the signal, and in Section 8.2, we describe its temporal behavior. In Section 8.3, we provide evidence for the signal being horizontally polarized. In Section 8.4, we describe the spectral behavior and show that it is consistent with the expected combination of Rayleigh scattering and emission from ice crystals. In Section 8.5, we show that the spatial scaling of power is consistent with the expectations of Kolmogorov turbulence. In Section 8.6, we show how signals from polarized atmosphere and water vapor scale with observing elevation. Lastly, we measure the anisotropic polarized atmospheric power amplitude for all observations and use these to create CDFs in Section 8.7.

**Table 1**
Percentile Values from the Measured $BT_\nu^2$ CDF and Their Comparison with the Results from Table 2 in R. S. Bussmann et al. (2005)

| Percentile (This Work) | 25 | 50 | 75 |
|---|---|---|---|
| $BT_{95}^2$ [mK$^2$rad$^{-5/3}$] | 0.68 | 1.5 | 3.3 |
| $BT_{150}^2$ | 4.5 | 11. | 28. |
| $BT_{220}^2$ | 16. | 43. | 130 |
| Percentile (Bussmann) | 27 | 54 | 81 |
| $BT_{150}^2$ | 3.7 | 10. | 37. |
| $BT_{220}^2$ | 11. | 38. | 160 |

**Note.** The SPT-3G results are upper limits because we have neglected the effect of wind speed in their calculation. The Bussmann values represent the 27th, 54th, and 81st percentiles of the CDF from which 7% of the data was removed due to producing bad fits to the model. The amplitudes corresponding to these missing data were assumed to be very high, and the reported CDF values can be taken as upper limits to the true 25th, 50th, and 75th percentiles.

### 8.1. Excess Noise in Q Polarization

We observe a strong asymmetry in the noise between the $Q$ (horizontal and vertical) and $U$ (+/−45°) polarization states, as can be clearly seen in Figure 13. This is a map produced from 2 hr of data with very high atmospheric polarization noise. This map has been processed identically to the maps used in the SPT-3G low-$\ell$ B-mode analysis. It is clear that the polarized atmosphere can be a very significant source of noise.

Elevated noise in the $Q$ timestreams (as seen in Figure 5) manifests as a significant excess in the ratio of the measured power spectra QQ/UU. This is true for both maps and 1D scans. The noise amplitude from all sources other than the sky is expected to be identical for both $Q$ and $U$. In particular, leakage of $T$ into $Q$ and $U$ should be similar and, based on our knowledge of detector gain matching, much smaller than the observed $Q$ signal. In the SPT-3G data, a high QQ/UU ratio can exist in the absence of a high TT signal, eliminating temperature to polarization leakage as the dominant source of excess $Q$ polarization noise.

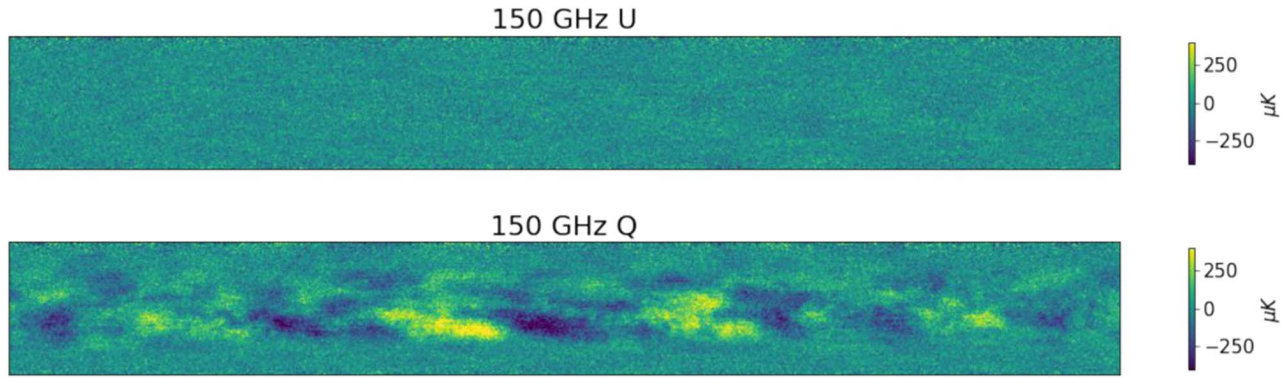

**Figure 13.** A ~66° × 8° map of a 2 hr SPT-3G observation that highlights the time-variable excess noise in $Q$. The maps for $Q$ (bottom) and $U$ (top) have the same color scale. While the $U$ map is largely featureless, the $Q$ map has large-scale noise for portions of the observation. A 10th-order polynomial is removed from the time-ordered data (TOD) used to produce this map. This filter removes signals with scales larger than ~10° in azimuth, including any DC component, and is chosen to optimize sensitivity to low-$\ell$ B-modes (which peak at $\ell \sim 100$). This is a higher-order filter than that applied to the $Q/U$ timestreams used in most of this work, which only have a linear drift removed.

The $Q$ polarization power is highly variable in time, while the $U$ noise is approximately constant and can generally be attributed to instrument noise. When characterizing properties of the signal such as the polarization angle, frequency, and spatial scaling, we select observations where the $Q$ noise power is significantly higher than the $U$ noise power. We restrict our analysis to the lowest subfield at $\delta = -44°.75$, as polarized atmosphere appears most strongly at low elevation. We define a set of "high-Q" observations, approximately 16% of the observations of the $\delta = -44°.75$ subfield, which we will use in the analysis described in Sections 8.3, 8.4, and 8.5. For an observation to be categorized as "high-Q," we require that the scan-averaged 220 GHz QQ power in the observation summed over the power spectrum bins corresponding to $15 < \ell < 100$ be greater than $5.9 \times 10^{-3}$ mK$^2$ rad. This cut corresponds to the high end of the UU distribution where QQ and UU noise diverge. In general, excess $Q$ polarization is not detected significantly in the 95 GHz band; therefore, in the following Sections, only 150 and 220 GHz data are used to characterize the polarized atmospheric signal. Note that these "high-Q" observations are not all of the observations where polarized atmosphere is detectable, but the fraction where we are confident polarized atmosphere dominates the large angular scale $Q$ noise power.

### 8.2. Time Dependence of Polarized Atmosphere

S. Takakura et al. (2019) found that periods of elevated polarized atmospheric signal last for ~30 minutes at the Chajnantor site in the Atacama Desert in Chile. We see similar timescales in our South Pole data, with elevated noise levels lasting anywhere from ~10 minutes to an entire 2 hr observation. Figure 14 shows an observation with QQ noise power varying strongly with time, consistent with the expected contribution from clouds of ice crystals. Figure 13 shows the impact of a large time-varying $Q$ signal on a map. There are two important timescales to note: scan duration and observation duration. As was shown in Section 6.2, the polarization anisotropy can be treated as a 2D screen being slowly moved by the wind. In general, the scan speed is sufficient that the structure can be approximated as being fixed ("frozen") on the sky. However, given the large (100°) azimuth scans of the telescope, by the time the telescope returns to the same spot, the sky will have changed sufficiently to be a nearly independent realization of the signal on scales up to several degrees. In this way, the signal from anisotropic atmospheric polarization can be averaged down in time. If the telescope scan speed is fast enough so that the atmospheric polarization anisotropy is not measured with high signal-to-noise in a single scan, then the atmosphere will not significantly impact the noise in the resulting observation maps. However, due to practical limitations on telescope scan speed, sensitive CMB experiments at the South Pole will generally detect the polarized atmosphere on large scales in some fraction of the data.

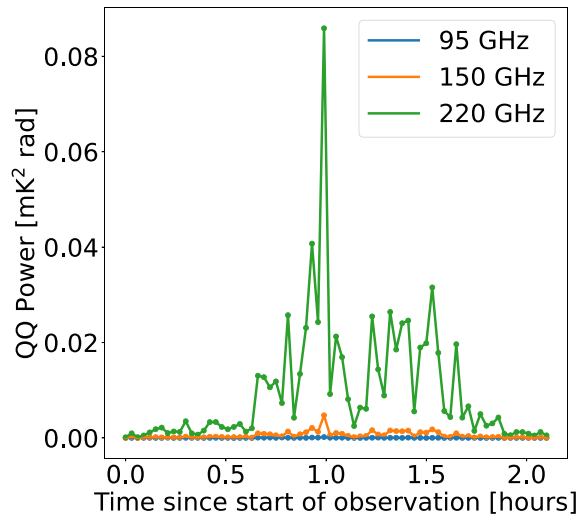

**Figure 14.** QQ power on large angular scales ($15 < \ell < 100$) at 95, 150, and 220 GHz for each scan over the course of a ~2 hr observation of the $\delta = -44°.75$ subfield. This is an extreme example that demonstrates both the temporal variability and frequency dependence of the signal.

### 8.3. Polarization Angle

Nonspherical ice crystals falling in the atmosphere will have their largest dimension aligned to be horizontal (K. Gustavsson et al. 2021). The theory laid out in Section 3.2 predicts that both scattering and thermal emission from these crystals will be horizontally polarized. Pure horizontal polarization will result in negative $Q$ and zero $U$ Stokes parameters. The CLASS experiment at the Chajnantor site (Y. Li et al. 2023) recently published measurements of polarization angle for brief periods of intense polarized emission. They accomplished this by measuring $Q$ and $U$ deviations from a constant baseline when visible clouds were present. However, we want to characterize the atmosphere under all conditions, particularly when fluctuations are small. The $Q$ timestreams we measure from each wafer are mean subtracted and, as is seen in Figure 5, display both positive and negative fluctuations about this mean. From these data alone, it is not possible to determine the polarization angle. We postulate that the fluctuations in polarization and temperature should be at least partially correlated. For the temperature fluctuations produced by ice, we know that this must be the case. Therefore, we compute the cross-correlation between $Q \times T$ and $U \times T$ for each observation, which recovers the correlated component of each. In the case of horizontal polarization from ice, we expect an increase in $T$ to be correlated with a decrease in $Q$ ($Q$ becomes more negative). We can then determine the polarization angle of the polarized fluctuations as

$$\psi = \frac{1}{2} \arctan\left(\frac{UT}{QT}\right). \quad (36)$$

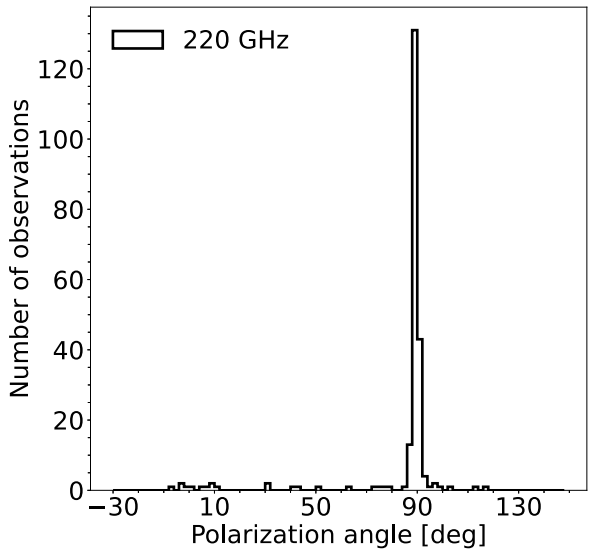

**Figure 15.** Polarization angle computed from $Q \times T$ and $U \times T$ for high-$Q$ observations. The large peak at $\psi = +90°$ corresponds to the horizontal polarization expected from scattering by horizontally aligned ice crystals.

The distribution of these polarization angles for the subset of "high-$Q$" observations defined in Section 8.1 is shown in Figure 15. We find that the polarization angle is strongly peaked at $\psi = +90°$, which corresponds to the expected horizontally polarized signal. It is worth noting that there is not a significant subpopulation of observations with $\psi = 0$ (vertical polarization), as was found in the analysis of atmospheric polarization from the CLASS experiment (Y. Li et al. 2023). This is likely due to the fact that the scattering at the South Pole during periods of strong polarized emission is dominated by horizontally aligned ice crystals, while the scattering observed by CLASS at the Chajnantor site may sometimes be dominated by spherical water droplets, resulting in periods of vertical polarization (Y. Li et al. 2023).

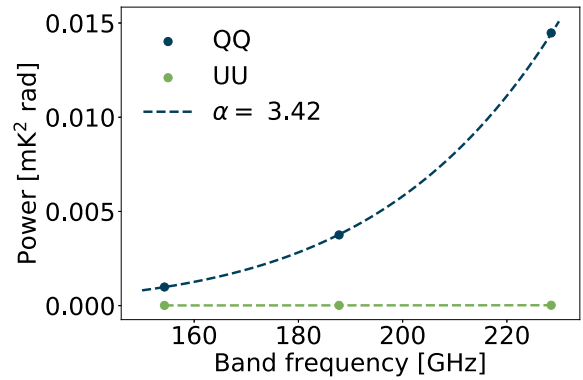

**Figure 16.** A sample 2 hr observation with high QQ power demonstrating a $P^{QQ}(\nu) \propto \nu^\alpha$ scaling consistent with polarized atmosphere. Cross spectrum power in each band is averaged over the range $15 < \ell < 100$, and UU is shown for comparison.

### 8.4. Frequency Dependence of Polarized Atmosphere

The multifrequency design of SPT-3G enables robust determination of the frequency scaling of the polarized atmospheric signal. A steep increase in the amplitude of the $Q$ signal with increasing frequency band is seen in Figure 14. We expect the QQ cross spectrum power to scale with frequency following a simple power law

$$P^{QQ} \propto ((\nu_i \nu_j)^{1/2})^{2\alpha}, \quad (37)$$

where $\nu_i$ and $\nu_j$ are the effective band centers for the $Q$ signals used in computing the cross spectra, and $\alpha$ is the spectral index that we wish to characterize.

To fit for the spectral scaling index, we first compute the average per-observation QQ cross spectrum power described in Section 5.1 over the range $15 < \ell < 100$ for the $150 \times 150$, $150 \times 220$, and $220 \times 220$ cross spectra. Due to the lower signal level, we omit the 95 GHz band and exclusively use "high-$Q$" observations, as defined in Section 8.1. We fit the data in log space and assume equal error bars across bands within an observation. The results of a fit to one observation with very high QQ power are shown in Figure 16.

The effective band centers of each frequency band (nominally 95, 150, and 220 GHz) depend on both the frequency response of the receiver and the frequency spectrum of the source we are measuring. As the source spectrum for polarized atmosphere is theorized to lie in the range from $\alpha = 2$ to 4, we calculate the effective frequency band centers iteratively. The source spectrum is first measured by fitting the frequency scaling of each observation-averaged cross spectrum using the nominal CMB band centers. The resulting source spectrum is then used to recalculate effective band centers, and this process is repeated until the source spectrum used to calculate the band centers and fitted frequency dependence converge.

In the distribution of measured spectral index values shown in Figure 17, the spectral index has a standard deviation of $\sigma_\alpha = 0.28$ and a mean value of $\alpha = 3.47 \pm 0.02$. The corrected band centers corresponding to this mean spectral index are 154.3 and 228.5 GHz for the nominal 150 and 220 GHz bands, respectively. The 150 x 220 GHz cross spectrum band center is given by $\sqrt{154.3 \times 228.5} = 187.8$ GHz.

The measured mean spectral index falls within the theoretical range expected from the combination of Rayleigh scattering and emission from ice crystals modeled in Section 3.2. The scatter about the mean is likely due to a combination of noise and variation in ice crystal properties. The mean spectral index is consistent with a dominant ice crystal equivalent radius of approximately $r = 100\ \mu$m. As described in Section 3, in the typical case of a distribution of ice crystal sizes, the observed signal and spectral index are dominated by the largest ice crystals. As described in Section 2.2.1, the preferred equivalent radius $r = 100\ \mu$m is similar to the upper end of the distribution of ice crystal diameters observed at the South Pole. Periods of significant atmospheric polarization are likely associated with the presence of these relatively rare large ice crystals.

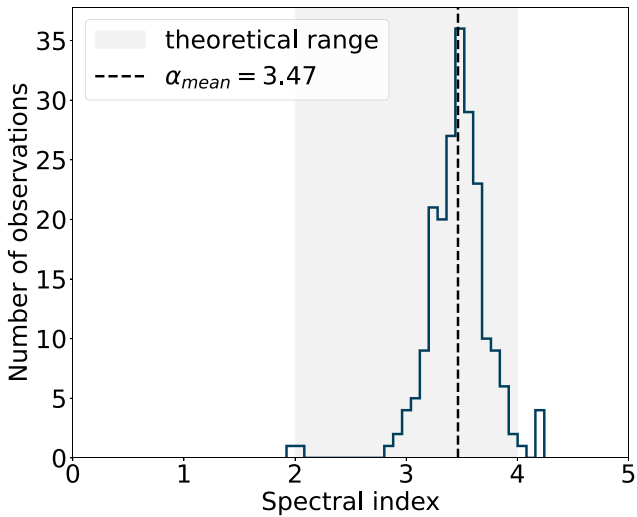

**Figure 17.** Distribution of QQ spectral indices for high-Q observations. The shaded region corresponds to the theoretical range predicted in Section 2.2.

These results can be contrasted with those of Y. Li et al. (2023), who used the CLASS experiment to measure a spectral index that was consistent with Rayleigh scattering between 90 and 150 GHz, and softened significantly by 220 GHz. They attribute this to either water vapor absorption or the presence of ice crystals large enough that Mie scattering is appropriate for the 220 GHz band. If such large crystals were present, it would imply a very large scattering amplitude per column density (IWP) of ice. This interpretation is consistent with the extremely large polarized signals measured by the POLARBEAR and CLASS experiments at the Chajnantor site.

To transform the amplitude measured in the 220 GHz band to that expected for other frequency bands, the amplitude is scaled by the measured power law from the effective band center $\nu = 228.5$ GHz,

$$BQ_\nu^2 = BQ_{220}^2 \left(\frac{\nu}{228.5\ \text{GHz}}\right)^{2\alpha}, \quad (38)$$

where $\alpha = 3.47$. This steep spectral scaling means that the power measured in the 150 and 95 GHz bands (effective band centers of 154.3 and 99.0 GHz) will be smaller by factors of 15.3 and 332, respectively. The very small polarized power in the 95 GHz band is the reason it is excluded from this analysis.

### 8.5. Spatial Scaling of Polarized Atmosphere

Here, we solve for the dependence of atmospheric polarization anisotropy power on spatial scale. We anticipate that the ice crystals responsible for the QQ fluctuation power are, like the water vapor fluctuations that dominate the temperature signal, passively entrained in the turbulence of the atmosphere. We expect that the QQ power measured in our 1D telescope scans will scale with angular frequency $\alpha$ (note $\alpha = \ell/2\pi$) as $P \propto \alpha^{b_{QQ}}$, where $b_{QQ} = -8/3$ for the expected Kolmogorov turbulence of the atmosphere. As with the spectral scaling, we carry out the spatial scaling portion of this analysis using only "high-Q" observations as defined in Section 8.1.

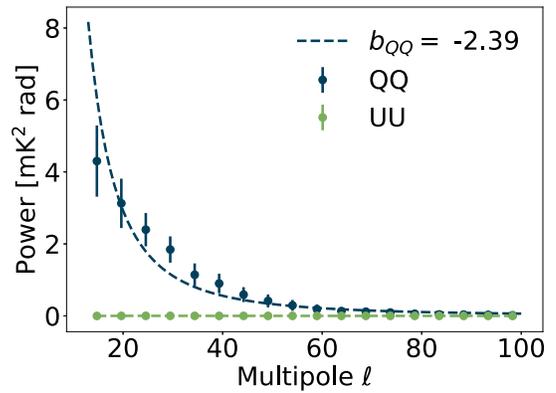

**Figure 18.** Spatial scaling of a sample high-Q observation with UU shown for comparison. Data points from the cross spectra are plotted with the best-fit power law.

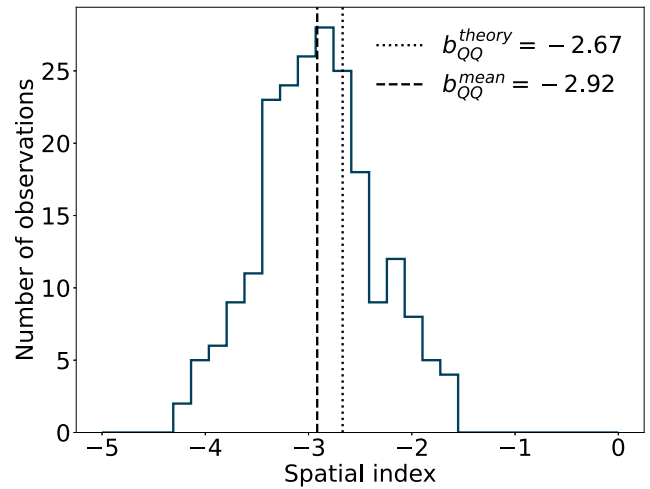

**Figure 19.** Distribution of QQ spatial scaling indices from high-Q observations. The dashed line marks the mean of the distribution, and the dotted line shows the theoretical expectation from Kolmogorov turbulence.

For each observation, we fit the 220 GHz cross spectra described in Section 5.1 to a power law. The data are fit using nonlinear least-squares over the range $15 < \ell < 100$. An example observation and fit is shown in Figure 18. The resultant histogram of spatial scaling indices, shown in Figure 19, has a standard deviation of $\sigma_b = 0.55$. The mean value is found to be $b_{QQ} = -2.92 \pm 0.04$. This spatial scaling is within 10% of both that found for temperature fluctuation power in Section 7.1 and the predictions of Kolmogorov turbulence.

Using the same procedure as for temperature fluctuations, we create a high signal average QQ power spectrum from the average of many observations. Cross spectra are selected by taking the subset of high polarized atmosphere observations used in Figure 19 and discarding the top 20% to avoid the average being dominated by a small number of extremely high-amplitude observations. QQ cross spectra from this set of observations are averaged and fit to a power law in log space, shown in Figure 20. Error bars for each $\ell$ bin in the summed power spectrum are calculated by combining the errors for that bin from each observation in quadrature and dividing by the number of observations. The final error bars in the plot are scaled to produce a reduced $\chi^2 = 1$ for the fit. As with the temperature power spectrum, the spatial scaling is visually consistent with a power law. The power-law fit yields a best-fit index of $b_{QQ} = -2.76 \pm 0.02$, within 5% of the Kolmogorov turbulence prediction.

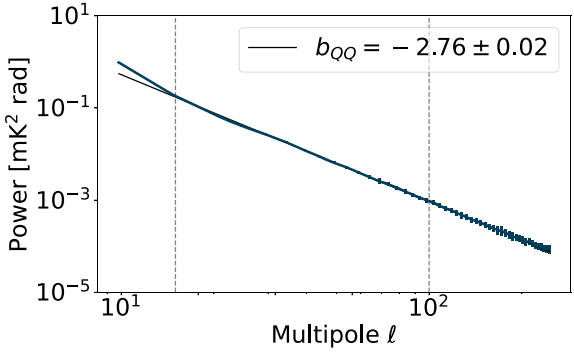

**Figure 20.** Spatial scaling of the average of high-$Q$ observation cross spectra. The vertical gray dotted lines denote fit range $15 < \ell < 100$. As is the case for TT power, the spatial scaling is well described by a power law that is roughly consistent with the prediction of Kolmogorov turbulence.

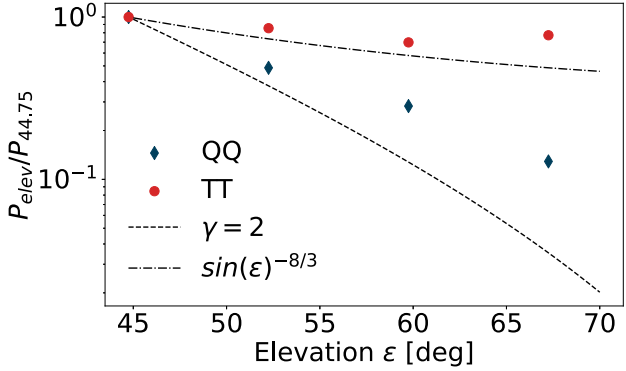

**Figure 21.** Scaling of measured QQ and TT power at 220 GHz with the elevation of the subfield. The excess in both TT and QQ at high elevation is consistent with the expected positive bias in power due to the measured median wind speed and the decreasing scan speed with increasing elevation.

### 8.6. Elevation Dependence of Polarized Atmosphere

In Section 4, theoretical expectations for the elevation dependence of both the QQ and TT power were presented. The TT elevation dependence is solely dependent on the geometry between the telescope elevation angle for a given observation and the layer of water vapor in the atmosphere. The QQ signal has an equivalent dependence on this geometry, but there is an additional dependence on elevation due to the orientation of the ice crystals themselves (described in Section 3.2). Both hexagonal platelets and columns will become aligned by gravity to create a horizontally polarized signal that decreases rapidly with increasing elevation angle. It is worth noting that the derivation of the elevation dependence of the polarization assumes that column-like crystals are randomly oriented in azimuth about the gravitational vector. In this limit, the polarization of the radiation from ice crystals vanishes for observations at zenith. However, if these crystals were preferentially aligned in azimuth angle by wind, then the polarization at high elevation would be larger, and the decrease with increasing elevation would be less steep.

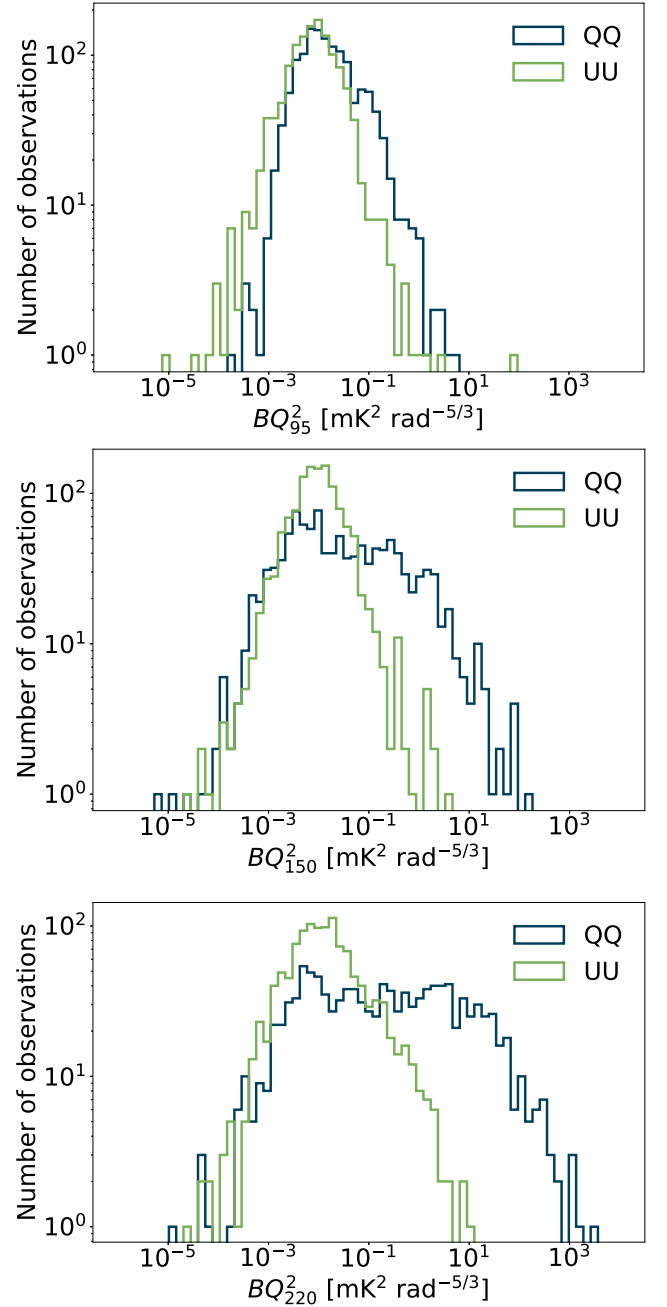

**Figure 22.** Distribution of positive $BQ_\nu^2$ and $BU_\nu^2$ values for 95 (top), 150 (middle), and 220 (bottom) GHz. The number of scans with anisotropy power (QQ) in excess of the expectation from instrument noise alone (UU) can be seen to increase steeply with observing frequency.

In Figure 21, we show the measured elevation scaling of TT and QQ atmospheric power and compare it with the theoretical predictions. The TT data points are calculated using the median cross spectrum power of all TT observations within each subfield over the range $15 < \ell < 100$. Given the highly variable nature of the polarization signal, we detect significant QQ power in only a fraction of observations. However, it is

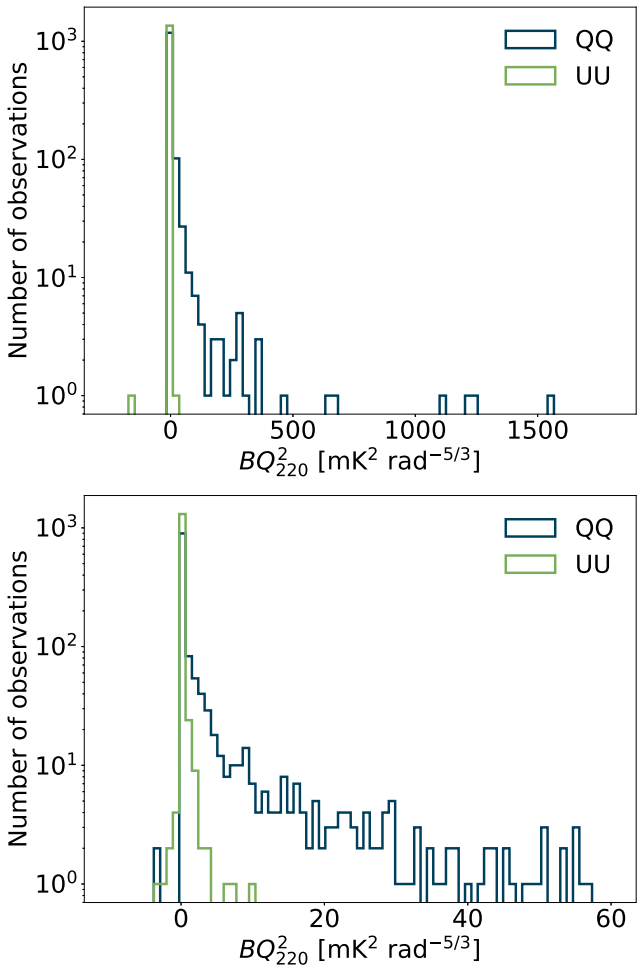

**Figure 23.** Histogram of $BQ^2_{220}$ amplitudes with a linear horizontal axis. The top panel shows the full distribution, while the lower panel shows a zoom-in on the lower-amplitude portion of the distribution.

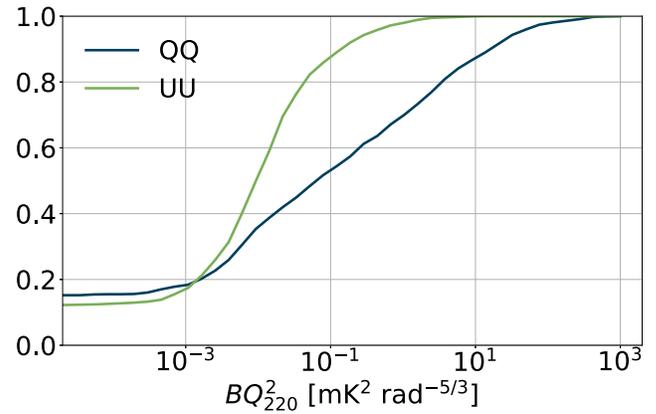

**Figure 24.** The cumulative distribution of $BQ^2_{220}$ and $BU^2_{220}$ for the $\delta = -44\overset{\circ}{.}75$ subfield.

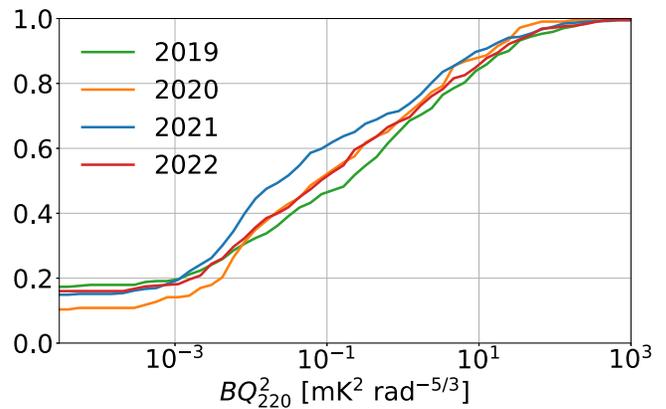

**Figure 25.** CDFs of observation QQ power amplitudes in the 220 GHz band for the $\delta = -44\overset{\circ}{.}75$ subfield for each Austral winter observing season.

**Table 2**
Percentile Values from the Cumulative Distribution Function of QQ Power Amplitudes for Each Observation

| Percentile | $BQ^2_{220}$ (mK$^2$ rad$^{-5/3}$) |
|---|---|
| 25 | $3.5 \times 10^{-3}$ |
| 50 | $6.4 \times 10^{-2}$ |
| 75 | 2.0 |

**Note.** A more complete tabulation of this CDF is included in Appendix B. For ~25% of scans, we detect no significant polarized power and, for the purpose of simulation, one can use the 25th percentile value as an upper limit to the true power for those observations.

reasonable to assume that the properties of the atmosphere for the subset of high QQ power scans will be similar for each subfield. The QQ data points are calculated using the median power of the highest 15% of observations in each subfield. This restricts our measurement to the regime of significant polarized atmosphere detection in all subfields. Figure 21 confirms the expected decline of atmospheric power with increasing observing elevation. We measure median QQ power to be 7.8 times higher for the $\delta = -44\overset{\circ}{.}75$ subfield than for the $\delta = -67\overset{\circ}{.}25$ subfield.

The excess power compared to the model with increasing elevation in both the TT and QQ data is likely the result of the increasing violation of the frozen sky approximation, as the telescope scans with a slower speed at higher elevation. Therefore, at high elevation angle, the measured amplitudes will be overestimated by a factor greater than that predicted in Section 6.2. In the limit where the frozen sky approximation is violated, the measured amplitude will depend on wind speed. The elevation angle analysis for QQ is based on the highest-amplitude scans, which are correlated with high wind speed due to this overestimation effect. Therefore, the overestimate of true sky power due to decreasing scan speed with increasing elevation will be greater for the QQ data. Despite these complications, the observed steep scaling of the observed QQ power with elevation is notable.

### 8.7. Amplitude Histogram

In this Section, we measure the amplitude of the polarized fluctuation power for every 2 hr observation over four winter observing seasons. These measurements of power amplitude are presented in the form of histograms and CDFs. We restrict the CDFs to the highest frequency band, 220 GHz, and the $\delta = -44\overset{\circ}{.}75$ subfield where the polarized power is largest. These results can be used to simulate an upper limit to the impact of polarized atmosphere on a given instrument and scan strategy operating at the South Pole with the caveat that the observations have a duration of ~2 hr.

In Figure 22, we present the distribution of positive $BQ_\nu^2$ values for each observation by band. The steep scaling with frequency is apparent when comparing QQ to the instrument noise-dominated UU distribution for each band. The full 220 GHz distribution is shown with a linear horizontal axis in Figure 23, including a zoom-in on the lower-amplitude portion of the distribution. In Figure 24, one can see that the UU cumulative distribution is only ∼12% at zero amplitude, indicating that UU has a slight bias toward positive power. In the absence of correlated noise, one would expect the UU distribution to be symmetric around zero. This positive bias could be caused by temperature to polarization leakage or other systematic effects; however, it is small enough that it does not affect the results of this work. The difference between the QQ and UU curves is the result of polarized atmosphere. The QQ power is above the 95th percentile in UU in 38% of observations, giving a rough estimate of how often detectable polarized atmosphere is present. In Table 2, we present CDF quartiles for the QQ 220 GHz data power amplitudes. A more detailed table of $BQ_{220}^2$ percentiles, drawn from the CDF in Figure 24, is given in Appendix B. We present $BQ_{220}^2$ CDFs by year in Figure 25 and show that the amplitude distribution of QQ power is relatively constant on a year-to-year basis at the South Pole.

To be sensitive to polarized atmospheric emission, an experiment needs not only high sensitivity, but excellent rejection of temperature to polarization leakage. It is worth noting that we are measuring extremely small values of QQ power compared to TT power. In the 220 GHz band and $\delta = -44°.75$ subfield, the average ratio of $P^{QQ}/P^{TT} = 6.5 \times 10^{-4} \pm 7.3 \times 10^{-6}$. This ratio is dependent on the QQ polarization fraction and therefore falls steeply with increasing elevation. Given the steeper frequency spectrum of scattering from ice than emission from water vapor, this ratio is expected to be a factor of ∼3.8 and ∼12.5 smaller for the 150 and 95 GHz bands, respectively. Careful control of temperature to polarization leakage is needed so that leaked temperature fluctuations are smaller than atmospheric polarization power and instrument noise. In the case of SPT-3G, we achieve this through precise gain matching of detectors with orthogonal polarization sensitivity.

As previously mentioned, the aim of reporting the CDF of $BQ_{220}^2$ values is to present a quantity that characterizes the atmosphere independent of observing elevation. These amplitudes are relevant for experiments with observations of 2 hr or longer, and intrinsically factor in time-correlation of the atmosphere on a 2 hr timescale. These results can be used to compare the atmosphere above the South Pole with the atmospheric conditions at other observing sites and simulate the impact of polarized atmosphere on planned experiments at the South Pole.

Without a full characterization of the amplitude distribution at the Chajnantor site, comparisons are limited to the extreme tail of high-amplitude scans reported by experiments operating there. Some of the polarized atmospheric signals reported by CLASS (Y. Li et al. 2023) at 220 GHz exceed $\Delta|Q| > 1\,\text{K}_{\text{RJ}}$. Similarly, the POLARBEAR experiment has reported polarized atmospheric signals with $\Delta|Q| > 0.3\,\text{K}_{\text{RJ}}$ at 150 GHz (S. Takakura et al. 2019). These signals correspond to polarized power that exceeds the largest-amplitude single scans measured in the 4 yr of Austral winter observations at the South Pole presented here. That being said, it is not particularly informative to compare South Pole Austral winter conditions with the extreme tail of observations with CLASS or POLARBEAR. Quantitative comparisons will require a similar statistical analysis of the conditions at the Chajnantor site.

## 9. Simulation of Signals from the Atmosphere

In order to simulate the impact of temperature and polarization fluctuations on maps produced by CMB telescopes, it is useful to generate realizations of the atmospheric temperature and polarization fluctuations. Here we provide a prescription for computing the equivalent CMB power spherical harmonic coefficients from the measured 2D atmospheric power spectrum. This can be used with standard CMB simulation tools to make an instantaneous full-sky realization of the atmospheric temperature or polarization fluctuations.

As discussed in Section 4, under certain assumptions, the 2D angular power spectrum of the instantaneous combined atmospheric emission and scattering signal can be expressed as

$$P(\alpha_x, \alpha_y) = B_\nu^2 f(\epsilon)(\alpha_x^2 + \alpha_y^2)^{-\beta/2}. \quad (39)$$

Here we have written the amplitude generally as $B_\nu^2$, which can be specified as corresponding to either temperature or polarization. In the case of TT power, the dependence on observing elevation, $f(\epsilon) = \sin(\epsilon)^{1-\beta}$, is just the geometrical factor described in Section 4. However, in the case of QQ power, the dependence on observing elevation needs to include the square of the polarization fraction and is given by Equation (21). The CDF percentile values of $BT_\nu^2$ are listed in Appendix A for all three SPT-3G observing bands. To scale these amplitudes to different frequencies, one would need to compute the ratio of differential emission from water vapor between the SPT-3G bands and the new frequency. The CDF percentile values of $BQ_{220}^2$ are listed in Appendix B and need to be scaled for other observing bands by Equation (37).

Assuming azimuthal symmetry, as in Section 4, we can write

$$P(\alpha) = B_\nu^2 f(\epsilon)(\alpha)^{-\beta}, \quad (40)$$

where $\alpha = \sqrt{\alpha_x^2 + \alpha_y^2}$. For a small patch of sky, we can approximate $\ell = 2\pi\alpha$ and express the flat-sky power spectrum as an angular power spectrum as a function of multipole number:

$$C_\ell \approx P(\ell) = B_\nu^2 f(\epsilon)\left(\frac{\ell}{2\pi}\right)^{-\beta}. \quad (41)$$

Adding a frequency-dependent factor $(dT_{\text{CMB}}/dT_{\text{RJ}})_\nu$ to put the RJ temperature of the atmospheric power in CMB temperature units, we have

$$C_\ell \approx B_\nu^2 \left(\frac{dT_{\text{CMB}}}{dT_{\text{RJ}}}\right)_\nu^2 f(\epsilon)\left(\frac{\ell}{2\pi}\right)^{-\beta}. \quad (42)$$

These spherical harmonic coefficients can be used to create full-sky realizations of the atmospheric temperature and polarization signal. It is worth noting, however, that these simulations will reflect an instantaneous snapshot of the sky and not the power in a map produced by an experiment that maps the sky over a finite period of time while the atmosphere drifts and changes. The polarized pattern on the sky will move and change over the course of an observation. However, the final full-season map will be a weighted average, or "coadd," of all of the observation-by-observation maps. Thus, this signal will average down in the final map. The impact on the final coadded map will depend in detail on the wind speed and instrument scanning pattern. A realization of the sky should have an amplitude drawn from the appropriate TT or QQ CDF, be moved with the median wind speed, and then mock observed with the specific instrument and scan strategy. This procedure will not reflect variations in sky power over the course of an observation, but should produce a simulated map with the appropriate power from the polarized atmosphere, particularly when used to simulate the result of many coadded maps.

## 10. Mitigation of Polarized Atmosphere for Power Spectrum Analysis

In this Section, we discuss techniques that can be used to mitigate the impact of polarized atmosphere and other sources of low-$\ell$/large angular scale noise. The techniques we describe below were developed for a forthcoming SPT-3G low-$\ell$ B-mode power spectrum analysis using data from the 2019 and 2020 observing seasons.

As described in Section 5, SPT-3G data is taken in 2 hr observations for each of the four subfields. The data from each 2 hr observation is then binned into a subfield map, and it is those subfield maps that are all combined to make one full-field full-depth map for cosmological analysis. In this analysis, we combine 3036 subfield observations from the 2019 and 2020 observing seasons into final 95, 150, and 220 GHz full-depth temperature and polarization maps.

Each one of the 3036 subfield $Q$ polarization maps contains four main components:

1. CMB signal: This signal is unchanging observation-to-observation.
2. Astrophysical foregrounds and other non-CMB signals, including galactic dust: This signal is unchanging observation-to-observation.
3. Detector and readout noise: This signal is different in each observation. It is the dominant source of noise at high-$\ell$/small angular scales and is a significant source of noise at low-$\ell$/large angular scales.
4. Polarized atmosphere: This signal is highly variable and different in each observation. This noise source is more significant at low-$\ell$/large angular scales.

The first two items in this list are the signal for which we want to maximize sensitivity in our full-depth maps. The latter two are noise terms. In the low-$\ell$ regime, polarized atmospheric noise is highly variable with the potential to be the dominant source of noise. We address these sources of noise by either subtracting them from the data in an unbiased way, and/or mitigating the impact of the excess noise through down-weighting or data cuts.

In Section 10.1, we describe using the spectral dependence of the polarized atmospheric signal to remove it from the 150 GHz band map for each observation. The atmospheric signal is highly variable in time, and removing it can significantly improve the low-frequency map noise.

Then, in Section 10.2, we describe combining the maps from individual observations using weights based on their low-frequency map noise. This optimizes the noise properties of the combined maps for low angular frequencies where inflationary B-modes may be detectable.

As will be described, the combination of these two techniques has the benefit of reducing the large angular scale atmospheric noise in the final coadded map without introducing bias and allows for optimal weighting of the cleaned observations in final full-depth maps.

### 10.1. Observation-by-observation Polarized Atmosphere Subtraction

The polarized atmosphere is highly variable, resulting in some maps with no significant signal from atmospheric polarization and some that are highly contaminated. With many observations, it is possible to separate the per-observation atmospheric signal and detector noise contribution to each map from the static CMB. Removing the polarized atmospheric signal from the individual observation maps reduces their large-scale noise. This avoids the loss in effective data volume that would result from simply cutting or downweighting the entire contaminated observation. The steep scaling of the atmospheric polarization power with frequency means that, for SPT-3G, this signal only contributes significantly to the 150 and 220 GHz bands. For each observation, we use the 220 GHz band to measure the atmospheric polarization signal and then use the known spectral scaling to remove it from the 150 GHz band data. For this work, we choose to demonstrate atmospheric cleaning on the 150 GHz band because it has higher signal-to-noise for the CMB signal than the 220 GHz band and is much more impacted by polarized atmosphere than the 95 GHz band. In this Section, we describe how we use the single observation 220 GHz maps to measure and remove the atmospheric polarization signal from the single observation 150 GHz maps.

#### 10.1.1. Processing the 220 GHz Map to Isolate Polarized Atmosphere

As mentioned above, a 220 GHz map of one observation contains ~4 distinct signals—the CMB, astrophysical foregrounds, detector noise, and polarized atmosphere. In the rest of this Section, we detail how to isolate the time-varying polarized atmosphere.

Removing CMB and Astrophysical Foregrounds: To remove CMB and astrophysical foregrounds, a 220 GHz full-depth full-field "coadd" is created. This is a weighted average of all of the observations from the 2019 and 2020 SPT-3G observing seasons. The weighting is done on a per-detector basis using the inverse of the variance of polarized noise power between 0.1 and 1 Hz. The result is a high signal-to-noise map of the 220 GHz sky, which is dominated by the CMB and astrophysical foregrounds. To remove these constant signals, this map is subtracted from the 220 GHz map for each observation. The resulting difference map then contains a realization of the polarized atmosphere and detector noise for the observation.

Filtering Detector Noise: Next, the 220 GHz map of polarized atmosphere and detector noise has a low-pass Butterworth filter applied at the angular scale where the map noise is detector-noise-dominated. This filter reduces the approximately white high-$\ell$ 220 GHz detector noise. Without this filtering, subtracting the 220 GHz map would degrade the small angular scale noise in the 150 GHz map. The resulting processed 220 GHz map for each observation is dominated by the large-scale polarized atmospheric signal that we want to remove.

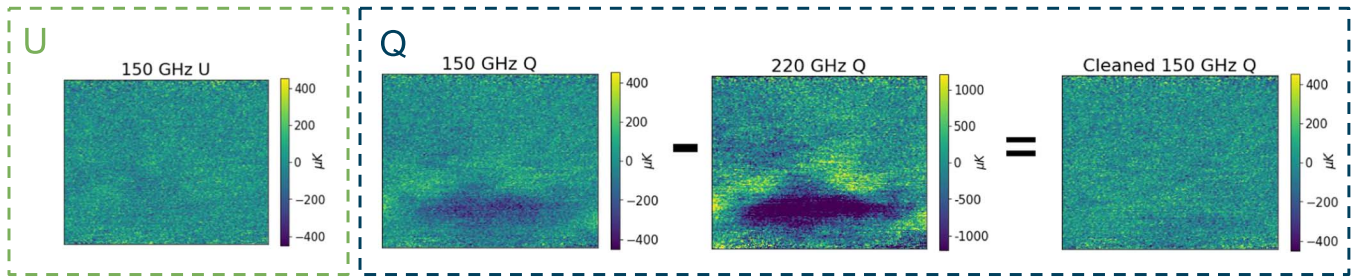

**Figure 26.** A graphic depiction of the band-dependence and $Q/U$ asymmetry of the polarized atmospheric signal. The left two plots are the 150 GHz $U$ (far left) and $Q$ (middle left) maps of a roughly $8° \times 8°$ cutout of a 2 hr SPT-3G observation filtered with an effective $30 < \ell < 3000$ bandpass, the same filtering as in Figure 13. The left two plots demonstrate the asymmetry between $Q$ and $U$ noise for an observation. The middle two plots show that the same spatial anisotropy exists in both the 150 GHz (middle-left) and 220 GHz (middle-right) map, but at different amplitudes as demonstrated by the difference in color bar. Subtracting a scaled copy of the 220 GHz map from the 150 GHz map shows the cleaned 150 GHz $Q$ map (far right), which now matches the noise levels of the 150 GHz $U$ map (far left).

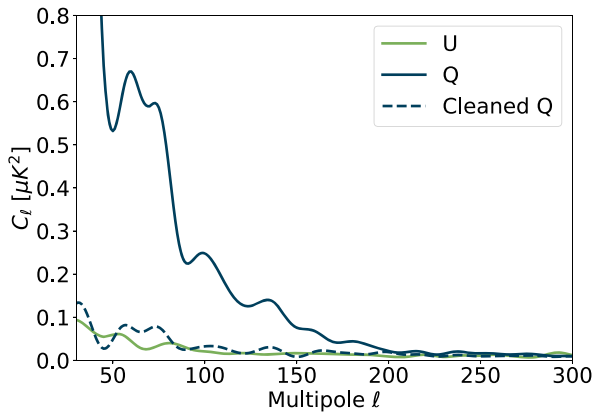

**Figure 27.** The effect of the 220 GHz subtraction on the 150 GHz $Q$ noise power spectrum for a map made from one 2 hr high-$Q$ power observation. The solid lines are $Q$ (blue) and $U$ (green) noise power spectra of one 2 hr observation. The blue dashed line is the $Q$ noise power spectrum after the polarized atmospheric signal has been reduced by cleaning with the 220 GHz data.

*10.1.2. Subtracting Polarized Atmosphere from the 150 GHz Data*

To subtract the polarized atmosphere within an observation, the processed 220 GHz map is scaled by a constant factor of 0.15. This map scaling factor was chosen to minimize the final coadded 150 GHz map noise. This technique takes advantage of the consistency of the frequency scaling of the polarized atmosphere, meaning the optimal 220 GHz map scaling factor will be approximately the same for every observation. This scaled 220 GHz $Q$ map is then subtracted from the 150 GHz $Q$ map. The scaling factor calculated in Section 8.4 from high-$Q$ amplitude scans of $\alpha = 3.47$ corresponds to a map scaling factor of 0.14. There is a negligible difference in the final 150 GHz $Q$ map noise using a map scaling factor of 0.14 versus 0.15.

The efficacy of this process can be seen in Figures 26 and 27. In Figure 27, the blue ($Q$) and green ($U$) solid lines correspond to the 150 GHz map power spectra from one highly contaminated observation before polarized atmosphere subtraction. At low-$\ell$, the polarized atmosphere results in roughly a factor of 20 more power in $Q$ than in $U$. The blue dashed line is the 150 GHz $Q$ power spectrum after this cleaning, which now has noise similar to that of the $U$ power spectrum (green). This is the quantitative Fourier representation of the reduction in noise seen in Figure 26.

This cleaning is performed on every 150 GHz map where the cleaned $Q$ power spectrum has less power than the baseline $Q$ power spectrum. In practice, this process reduces the noise for $\sim$70% of observations. We do not perform any atmosphere subtraction on the 95 GHz maps as the steep spectral scaling of the signal means that the polarized atmospheric power has little impact on the overall noise for the 95 GHz band.

Due to the mean sky subtraction and fixed frequency scaling, this algorithm is linear and does not impact the map transfer function, remove sky signal, or create bias. Signals fixed on the sky are unaffected since the full-depth 220 GHz coadded map (a high signal-to-noise map of the sky) is subtracted from each 220 GHz observation map before removing it from the 150 GHz map. This reduces low-frequency map noise without impacting constant signals such as the CMB and astrophysical foregrounds.

*10.2. Low-$\ell$ Weighting in Timestreams and Maps*

In addition to observation-by-observation polarized atmospheric subtraction, we make three choices for these maps that are different from other SPT-3G analyses (D. Dutcher et al. 2021; Z. Pan et al. 2023) to improve the large angular scale noise.

*Elevation Slew Gains.* The relative detector gains within a polarization pixel pair are determined by minimizing the response of the detector difference to an elevation slew, instead of the conventional method of matching the response between detectors to an unpolarized astrophysical source. The measurement of large angular scale polarization requires excellent gain matching between detectors. Mismatches in frequency bandpasses or beams between detectors can lead to temperature to polarization leakage when the gains are determined from sources that are not beam filling or have a different spectrum than the atmosphere. Using the elevation-slew-determined relative gains reduces the temperature to polarization leakage from atmospheric temperature fluctuations and results in an $\sim$20% improvement in large angular scale noise.

*Polarization Pixel-pair Low-frequency Weights.* The relative weighting of detectors in an observation map is determined by the inverse variance of 0.1–1 Hz power for timestreams created by subtracting orthogonal polarization pixel pairs. The pair subtraction removes the common-mode temperature signal so that the weighting reflects the noise in polarization. The pixel pair weights are determined using the noise in a low-frequency band that approximately corresponds to the spatial scales being targeted in the low-$\ell$ B-mode analysis. This step downweights the contributions of detectors with high low-frequency noise in the final map.

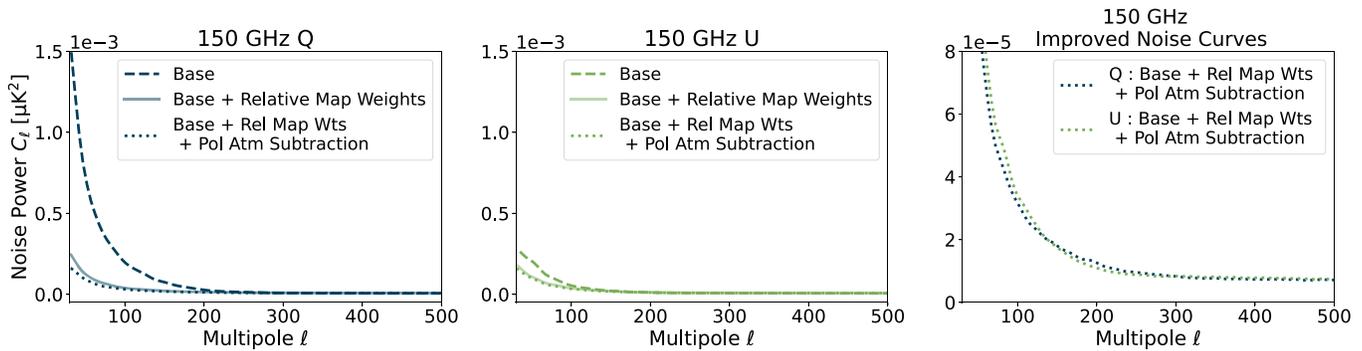

**Figure 28.** Noise power as a function of angular scale from 2 yr (2019–2020) of SPT-3G Austral winter observations for both $Q$ and $U$. The dashed line (labeled "Base" above) is calculated from the default low-$\ell$ optimized maps. They are made with a 10th-order polynomial TOD filter, gain matching using elevation slew response, and polarization pixel-pair low-frequency weights. The solid line corresponds to additionally weighting on correlated low-$\ell$ map noise (labeled "Relative Map Weights" above), as described in Section 10.1.1. The dotted line demonstrates the improvement if one completes all of the above, and in addition, completes an observation-by-observation polarized atmosphere subtraction (labeled "Pol Atm Subtraction" above). From the initial maps with the dashed curve, a 5x improvement in polarized noise power between $30 < \ell < 100$ can be gained from these targeted polarized atmosphere mitigation strategies. The far-right panel shows the final noise curves with all of the improvements, showcasing the equal $Q$ and $U$ noise.

*Weighting on Correlated Map Noise.* In the reduction of the data, a "weights map" is produced in conjunction with the individual observation data map. This map is constructed from the scan-by-scan sum of the inverse of the variance of polarized noise power between 0.1–1 Hz in the difference of detector pixel-pair timestreams, and used for relative weighting between maps. However, this weight will not accurately reflect the noise in the map, since it does not take into account atmospheric noise that is correlated between detectors. Once observation maps are made, these "weights maps" are normalized to the inverse of the map polarization power spectrum between $\ell = 50$ and 250. These new scaled weights reflect the low-frequency noise in the observation map, including correlated sources such as polarized atmosphere. The signal-to-noise for sources on the sky in each observation is so low that the bias incurred by weighting on the map power is negligible. For higher signal-to-noise maps, one would subtract the 150 GHz season mean map from each observation map before calculating the weights.

The relative efficacy of these mitigation techniques can be seen in Figure 28. The dashed line is the mean noise generated from 100 random "signflips" of two seasons of SPT-3G data—a process that separates all of the data into two equally weighted bundles, multiplies one bundle by $-1$, and coadds the bundles. This nulls the sky signal, but the noise properties are the same as the full two-season data coadd. This dashed line includes elevation slew gains and low-frequency inverse variance weighting on pixel pairs. The solid line shows the efficacy of additionally weighting observation maps on low-frequency map noise. The dotted line demonstrates the result using all of the above, and in addition, map-based polarized atmosphere subtraction. These analysis choices reduce the initial $Q$ noise power by a factor of 7.5 and the $U$ noise power by a factor of 2 over $30 < \ell < 100$, leading to roughly equivalent final $Q$ and $U$ noise levels, signifying a lack of polarized atmospheric contamination, as seen in the right panel of Figure 28. When $Q$ and $U$ are combined to produce E or B maps, polarized noise power in the SPT-3G coadded 150 GHz map is reduced by more than a factor of 5 over the same $\ell$ range.

The polarized atmosphere is highly variable with a long tail toward high-$Q$ polarization power. With SPT-3G, a fraction of observations see significant polarized atmospheric power on the angular scales being targeted for B-mode searches. By using low-$\ell$ map noise to normalize the map weights, we can downweight this tail of highly polarized maps, resulting in a significant improvement in the low-$\ell$ noise of the final coadded map. However, by downweighting observation maps based on their raw low-$\ell$ map noise, we lose their contribution to the total data volume. The map-based frequency subtraction described above reduces the polarized signal in the observation maps before they are weighted. This makes it possible to recover observations that would otherwise have been more highly downweighted and include them in the final coadd. In practice, the map-based frequency subtraction makes it possible to increase the effective data volume by $\sim$20% between $30 < \ell < 100$ over that achieved by weighting on the raw map noise.

The frequency subtraction described here takes advantage of a specific feature of the SPT-3G detector design—copointing multifrequency pixels. This property allows us to directly subtract different frequency bands. However, many existing or planned experiments such as the Simons Observatory (P. Ade et al. 2019) or CMB-S4 (K. Abazajian et al. 2022) have not baselined simultaneous coverage across a broad frequency range, which is required for this style of polarized atmosphere subtraction. At the current sensitivity of SPT-3G, map-based frequency subtraction only improves the low-$\ell$ noise power by $\sim$20% from that obtained through map weight normalization alone. This suggests that downweighting the relatively few observations with highly polarized atmosphere is, at least in the case of SPT-3G, an effective mitigation strategy.

It is important to note that the impact of atmospheric temperature and polarization fluctuations will be largest for signals with angular scales comparable to or larger than the field of view of the telescope. Atmospheric fluctuations on these scales are coherent across the focal plane and add constructively in the resulting map. For a large-aperture telescope such as SPT, with a field of view of $\sim$2°, the polarized atmosphere contributes nearly maximally to the degree angular scales where the primordial B-mode signal peaks. In contrast, small-aperture telescopes will be less affected by noise from the polarized atmosphere on the scales relevant for measurements of primordial B-modes. In the case of BICEP/Keck, with a field of view of ∼15°, there are many degree-scale atmospheric polarization fluctuations across the telescope field of view. When the data is combined in a map, the polarization fluctuations add incoherently and contribute less noise power than the same polarization fluctuations would for SPT. The quantitative impact of the polarized atmosphere on a given experiment will depend on the observing site, instrument sensitivity, and scan strategy.

## 11. Conclusions

Using observations with the SPT-3G CMB receiver on the 10 m diameter SPT, we have detected a highly variable polarized signal from the atmosphere. We present an analytic description of the polarized signal produced by ice crystals in the atmosphere. This signal contains contributions from Rayleigh scattering of thermal emission from the ground and thermal emission from the ice. The spatial distribution of ice crystals is anisotropic and, like the distribution of water vapor, follows a Kolmogorov power law. The main prediction of this model is a horizontally polarized signal manifesting as excess noise in Stokes $Q$ with a polarization angle of $+90°$. The polarized signal is also predicted to be a steep function of observing frequency and telescope elevation.

We have measured and characterized emission from the atmosphere above the South Pole during 4 yr of Austral winter observing. The SPT-3G receiver consists of 10 detector wafers each with ∼1600 polarization sensitive detectors equally distributed between three observing bands centered at 95, 150, and 220 GHz. Each of the detector wafers in the array is used to produce an instantaneous measurement of $I$ (or $T$), $Q$, and $U$ Stokes parameters. We verify that the temperature and polarization fluctuations move on the sky with wind at the same angular speed and that it is an acceptable approximation to consider the sky fluctuations as stationary for the SPT telescope scan speed. Cross correlations between signals from the different detector wafers are used to produce unbiased estimates of power in $T$, $Q$, and $U$. Wafer cross spectra are used to determine the spatial scaling, frequency scaling, and dependence on observing elevation angle of both the temperature and polarization fluctuation power.

We first verify our analysis method through comparisons of TT power with measurements made by R. S. Bussmann et al. (2005) and model predictions. The angular scaling of TT power is consistent with the predictions of Kolmogorov turbulence, and the frequency scaling is consistent with the emission being dominated by water vapor. Significant temperature anisotropy power is measured in every 2 hr subfield observation, and the distribution of measured amplitudes is consistent with previous measurements at the South Pole by R. S. Bussmann et al. (2005). As predicted by the geometry of the observation and the Kolmogorov spatial scaling, the observed TT power decreases slowly with increasing observing elevation, $\epsilon$, as $P_{TT} \propto \sin(\epsilon)^{-8/3}$.

We then present a complete characterization of the polarization signal. This signal manifests as excess large angular scale power in the QQ power spectrum, while the UU power is generally consistent with detector and readout noise. The spatial scaling of the QQ power spectrum is similar to that of the TT power spectrum and the predictions of Kolmogorov turbulence. The scaling of QQ power with frequency is very steep with $P_{QQ} \propto (\nu)^{2\alpha}$, where the spectral index $\alpha = 3.47 \pm 0.02$ is consistent with the combination of polarized scattering of radiation from the ground and thermal emission by a distribution of ice crystals with a maximum equivalent radius of ∼100 $\mu$m. The polarization angle of the fluctuations is found to be $\psi = +90°$, consistent with the expectation of the polarized signal arising from horizontally aligned ice crystals.

As predicted by the model, the measured QQ power falls steeply with increasing observation elevation angle. The QQ power measured at elevation $\epsilon = 44°.75$ is a factor of 7.8 higher than that observed at $\epsilon = 67°.25$. This decline is softened by bias in the high elevation angle measurements due to wind speed effects, significantly decreasing the measured ratio from the predicted ratio of 28.2.

The amplitude of the polarized signal is highly variable, and we only detect it in a fraction of observations, even at the highest observing frequency (220 GHz) and lowest-elevation (44°.75) subfield. We present histograms and CDFs of the measured 220 GHz QQ power amplitudes for every 2 hr observation of the $\delta = -44°.75$ subfield for four Austral winter observing seasons. These results can be easily scaled to other observing frequencies and observing elevations.

We show that, because the amplitude of the polarized atmospheric signal is so highly variable, the impact on the sensitivity of an experiment can be greatly reduced by downweighting or cutting the small fraction of observations that detect significant polarized atmosphere. In addition, we are able to make use of the consistent steep spectral scaling of the polarized signal to clean the 150 GHz SPT-3G maps by subtracting a scaled version of the 220 GHz maps. This makes it possible to recover the fraction of contaminated maps that would otherwise be downweighted. These combined techniques reduce the polarized noise power by a factor of 5 between $30 < \ell < 100$ in the SPT-3G data set.

Future experiments could be designed to mitigate the impact of the polarized atmosphere by observing over a broad frequency range to improve spectral subtraction, distributing detectors between several independent telescopes observing independent sky, and scanning telescopes faster so that fluctuations are imaged with lower signal-to-noise. However, we demonstrate that simply downweighting the fraction of observations with significant polarized atmosphere is an effective mitigation strategy for SPT-3G.

The results presented here can be used to simulate the impact of polarized atmosphere on millimeter-wavelength observations of the CMB at the South Pole for any combination of instrument design, observation strategy, and analysis choices. We anticipate that this will be particularly useful to current and planned experiments seeking to measure inflationary B-mode CMB polarization. We encourage similar quantitative studies of polarized atmospheric power at other CMB observing sites, in particular the Chajnantor site in the Atacama Desert, in order to facilitate detailed simulations and site comparisons.

## Acknowledgments

The South Pole Telescope program is supported by the National Science Foundation (NSF) through the award OPP-1852617. Partial support is also provided by the Kavli Institute of Cosmological Physics at the University of Chicago. A.C. was supported by the National Science Foundation Graduate Research Fellowship Program under grant No. DGE 1752814. Support for this work for J.Z. was provided by NASA through the NASA Hubble Fellowship grant HF2-51500 awarded by the Space Telescope Science Institute, which is operated by the Association of Universities for Research in Astronomy, Inc., for NASA, under contract NAS5-26555. S.T. was supported by JSPS Overseas Research Fellowship from the Japan Society for the Promotion of Science. Argonne National Laboratory's work was supported by the U.S. Department of Energy, Office of High Energy Physics, under contract DE-AC02-06CH11357. We would like to thank J.R. Lewis (University of Maryland Baltimore County) for the helpful comments and discussions about the MPLNET lidar at the South Pole.

## Appendix A
## $BT_\nu^2$ CDF

We present $BT_\nu^2$ CDF percentile values (Table 3) for use in modeling taken from the observation CDF shown in Figure 11.

**Table 3**
$BT_\nu^2$ CDF Upper Percentile Limits for the Three SPT-3G Frequency Bands

| Percentile | $BT_{95}^2$ (mK$^2$ rad$^{-5/3}$) | $BT_{150}^2$ | $BT_{220}^2$ |
|---|---|---|---|
| 5  | 0.30 | 1.5  | 4.9 |
| 10 | 0.37 | 2.2  | 7.2 |
| 15 | 0.47 | 2.9  | 9.8 |
| 20 | 0.57 | 3.6  | 12. |
| 25 | 0.68 | 4.5  | 16. |
| 30 | 0.80 | 5.6  | 20. |
| 35 | 0.94 | 6.6  | 24. |
| 40 | 1.1  | 8.1  | 29. |
| 45 | 1.3  | 9.4  | 35. |
| 50 | 1.5  | 11.  | 43. |
| 55 | 1.7  | 13.  | 53. |
| 60 | 2.0  | 16.  | 65. |
| 65 | 2.3  | 19.  | 79. |
| 70 | 2.7  | 23.  | 98. |
| 75 | 3.3  | 28.  | 130 |
| 80 | 4.1  | 36.  | 160 |
| 85 | 5.3  | 47.  | 210 |
| 90 | 8.5  | 70.  | 340 |
| 95 | 19.  | 140  | 740 |

## Appendix B
## $BQ_\nu^2$ CDF

We present $BQ_\nu^2$ CDF percentile values (Table 4) for use in modeling taken from the observation CDF shown in Figure 24.

**Table 4**
$BQ_{220}^2$ Upper Percentile Limits

| Percentile | $BQ_{220}^2$ (mK$^2$ rad$^{-5/3}$) |
|---|---|
| 25 | $3.5 \times 10^{-3}$ |
| 30 | $5.8 \times 10^{-3}$ |
| 35 | $9.0 \times 10^{-3}$ |
| 40 | $1.7 \times 10^{-2}$ |
| 45 | $3.5 \times 10^{-2}$ |
| 50 | $6.4 \times 10^{-2}$ |
| 55 | 0.14 |
| 60 | 0.26 |
| 65 | 0.56 |
| 70 | 1.1 |
| 75 | 2.0 |
| 80 | 3.6 |
| 85 | 7.5 |
| 90 | 16. |
| 95 | 42. |

**Note.** Below the 25th percentile, there is no significant detection of polarized atmosphere, and conservative estimates should assume the 25th percentile value for the rest of the distribution.

## ORCID iDs

A. Coerver ● https://orcid.org/0000-0002-2707-1672
A. J. Anderson ● https://orcid.org/0000-0002-4435-4623
M. Archipley ● https://orcid.org/0000-0002-0517-9842
L. Balkenhol ● https://orcid.org/0000-0001-6899-1873
A. N. Bender ● https://orcid.org/0000-0001-5868-0748
B. A. Benson ● https://orcid.org/0000-0002-5108-6823
F. Bianchini ● https://orcid.org/0000-0003-4847-3483
L. E. Bleem ● https://orcid.org/0000-0001-7665-5079
F. R. Bouchet ● https://orcid.org/0000-0002-8051-2924
T. W. Cecil ● https://orcid.org/0000-0002-7019-5056
P. M. Chichura ● https://orcid.org/0000-0002-5397-9035
T. M. Crawford ● https://orcid.org/0000-0001-9000-5013
C. Daley ● https://orcid.org/0000-0002-3760-2086
D. Dutcher ● https://orcid.org/0000-0002-9962-2058
K. R. Ferguson ● https://orcid.org/0000-0002-4928-8813
A. Foster ● https://orcid.org/0000-0002-7145-1824
R. Gualtieri ● https://orcid.org/0000-0003-4245-2315
G. P. Holder ● https://orcid.org/0000-0002-0463-6394
M. Millea ● https://orcid.org/0000-0001-7317-0551
G. I. Noble ● https://orcid.org/0000-0002-5254-243X
Z. Pan ● https://orcid.org/0000-0002-6164-9861
K. A. Phadke ● https://orcid.org/0000-0001-7946-557X
A. Rahlin ● https://orcid.org/0000-0003-3953-1776
C. L. Reichardt ● https://orcid.org/0000-0003-2226-9169
J. A. Sobrin ● https://orcid.org/0000-0001-6155-5315
C. Umilta ● https://orcid.org/0000-0002-6805-6188
N. Whitehorn ● https://orcid.org/0000-0002-3157-0407
W. L. K. Wu ● https://orcid.org/0000-0001-5411-6920